\documentclass{aa}

\usepackage{graphicx}
\usepackage{txfonts}
\usepackage{subcaption}
\usepackage{placeins}
\usepackage[colorlinks=true,
            linkcolor=blue,
            citecolor=blue,
            urlcolor=blue]{hyperref}
\usepackage{newtxtext,newtxmath}
\usepackage[T1]{fontenc}
\usepackage{amsmath}
\usepackage{amssymb}
\usepackage{color}
\usepackage{hyperref}
\usepackage{pdflscape}
\newcommand{\orcid}[1]{\href{https://orcid.org/#1}{\includegraphics[width=10pt]{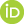}}}
\usepackage{lipsum} 
\usepackage{leftidx}
\usepackage{float}
\usepackage{pdflscape}
\usepackage{tabularx}
\usepackage[font=footnotesize]{caption}
\usepackage{multirow}
\definecolor{forestgreen}{rgb}{0.0, 0.7, 0.1}

\usepackage{rotating}
\newcommand{\nodata}{\ldots}

\usepackage[normalem]{ulem}
\newcommand{\revdel}[1]{\textcolor{red}{}}
\newcommand{\revadd}[1]{#1}

\usepackage{orcidlink}
\let\orcid\orcidlink

\begin{document}

\title{The local ultraviolet signature of Type Ia supernova environments from HST and MUSE}
\titlerunning{UV of SNe Ia hosts}
\authorrunning{Galbany et al.}

\author{
Llu\'is~Galbany\inst{1,2}\thanks{\email{l.g@csic.es}}\orcid{0000-0002-1296-6887},
Cullen Abelson\inst{3}\orcid{},
Alaa Alburai\inst{1,2}\orcid{},
Joseph P Anderson\inst{4}\orcid{},
Yago Ascasibar\inst{5,6}\orcid{},
Chris Ashall\inst{7}\orcid{},\\
\mbox{Carles Badenes\inst{3}\orcid{},}
Chris Burns\inst{8}\orcid{},
\mbox{\`Elia Di\'eguez Gurn\'es\inst{1}\orcid{},}
\mbox{Albert Garc\'ia Soto\inst{1}\orcid{},}
Claudia P. Guti\'errez\inst{1,2}\orcid{},\\
Eric Y. Hsiao\inst{9}\orcid{},
Hanindyo Kuncarayakti\inst{10,11}\orcid{},
Andrew J. Levan\inst{12,13}\orcid{},
\mbox{Jospeh Lyman\inst{13}\orcid{},}
Mark M. Phillips\inst{8}\orcid{},\\
\mbox{Sebastian F. S\'anchez\inst{14,15}\orcid{},}
Ramon Sanfeliu\inst{2,1}\orcid{},
Maximilian Stritzinger\inst{16}\orcid{},
Danny Steeghs\inst{13}\orcid{}
}

\institute{
Institute of Space Sciences (ICE-CSIC), Campus UAB, Carrer de Can Magrans, s/n, E-08193 Barcelona, Spain
\and
Institut d'Estudis Espacials de Catalunya (IEEC), 08860 Castelldefels (Barcelona), Spain
\and
Department of Physics and Astronomy and PITT PACC, University of Pittsburgh, Pittsburgh, PA 15260, USA
\and
European Southern Observatory, Alonso de C\'ordova 3107, Vitacura, Casilla 19001, Santiago, Chile
\and
Departamento de Física Te\'orica, Universidad Aut\'onoma de Madrid (UAM), Madrid 28049, Spain
\and
Centro de Investigación Avanzada en Física Fundamental (CIAFF-UAM), Madrid 28049, Spain
\and
Institute for Astronomy, University of Hawai‘i, Honolulu, HI 96822, USA
\and
Las Campanas Observatory, Carnegie Observatories, Casilla 601, La Serena, Chile
\and
Department of Physics, Florida State University, 77 Chieftan Way, Tallahassee, FL 32306, USA
\and
Finnish Centre for Astronomy with ESO (FINCA), 20014 University of Turku, Finland
\and
Tuorla Observatory, Department of Physics and Astronomy, 20014 University of Turku, Finland
\and
Department of Astrophysics/IMAPP, Radboud University, 6525 AJ Nijmegen, The Netherlands
\and
Department of Physics, University of Warwick, Coventry, CV4 7AL, UK
\and
Instituto de Astronom\'ia, Universidad Nacional Auton\'oma de M\'exico, A.P. 106, Ensenada 22800, BC, M\'exico
\and
Instituto de Astrof\'isica de Canarias, La Laguna, Tenerife, E-38200, Spain
\and
Department of Physics and Astronomy, Aarhus University, Ny Munkegade 120, DK-8000 Aarhus C, Denmark
}

\date{Received \today; accepted XXX}

\abstract
{
%Context: 
The local environment of a Type Ia supernova (SN~Ia) \revdel{encodes information on progenitor age} \revadd{traces the local stellar-population age (a statistical proxy for the progenitor age)}, star formation, dust, and chemical enrichment that may influence  both its light-curve properties and its standardized luminosity.
%Aims: 
We assemble a homogeneous explosion-site dataset to test how near-ultraviolet (UV) imaging combined with optical integral-field spectroscopy improves the characterization of SN~Ia environments and their connection to SN~Ia observables.
%Methods: 
We analyze 340 low-redshift SN~Ia explosion sites observed with HST/WFC3 in F275W together with matched VLT/MUSE spectroscopy. For each site we measure local photometry and extract a 1~kpc-aperture spectrum, model the stellar continuum with \texttt{STARLIGHT} in a joint UV+optical fit, derive local stellar and gas-phase properties, and fit the SN light curves with \texttt{sncosmo}/SALT3-NIR.
%Results: 
UV emission is detected at 235 of the 337 sites with valid HST imaging. The UV- and H$\alpha$-based star-formation-rate surface densities are strongly correlated, with the UV estimates systematically higher by about 0.5 dex. Including the UV constraint in the \texttt{STARLIGHT} fit shifts local luminosity-weighted ages toward older values by a median of $+0.20$ dex for UV detections and $+0.45$ dex for UV non-detections, showing that UV data help break age--dust--metallicity degeneracies and \revdel{improve local stellar-population age estimates} \revadd{substantially alter and tighten the inferred local stellar-population ages within the adopted modelling framework}. Consistently, F275W$-r$ is very strongly correlated with the joint-fit stellar age. Within the 169-SN~Ia calibration sample, the joint stellar age, the H$\alpha$ specific star-formation rate (sSFR), and the UV sSFR all correlate strongly with the SALT3 stretch parameter $x_1$, with significances above $6.8\sigma$. After standardization, the residual Hubble-residual trends are weaker: the strongest \revdel{signal is} \revadd{trend is} found for the local mass-weighted age, while the UV and H$\alpha$ sSFRs show residual offsets of comparable sign and amplitudes of order $\sim 0.05$ mag. \revadd{None of the environmental dependencies of the Hubble residual exceeds $3\sigma$}.
%Conclusions: 
The results \revdel{support a primary age--stretch mechanism} \revadd{point to a dominant empirical association between light-curve stretch and a covariant young-stellar-population axis}, largely absorbed by the standard light-curve corrections, while any residual luminosity dependence on the local environment is weaker and more subtle. If confirmed with larger samples, these residual trends may still become relevant for the younger SN~Ia populations expected at higher redshift, and should therefore be tested in cosmology analyses with forthcoming Roman and JWST SN samples.}
%{}{}{}{}

\maketitle
\nolinenumbers

%%%%%%%%%%%%%%%%%%%%%%%%%%%%%%%%%%%%%%%%%%%%%%%%%%%%%%%
%%%%%%%%%%%%%%%%%%%%%%%%%%%%%%%%%%%%%%%%%%%%%%%%%%%%%%%
%%%%%%%%%%%%%%%%%%%%%%%%%%%%%%%%%%%%%%%%%%%%%%%%%%%%%%%

\section{Introduction}\label{intro}

Type Ia supernovae (SNe Ia) are cornerstones of observational cosmology, providing the most direct and mature evidence for the accelerating expansion of the Universe and remaining among the most powerful probes of the dark energy equation of state \citep{1998AJ....116.1009R,1999ApJ...517..565P}. Their utility as standardisable candles rests on well-established empirical relations between peak luminosity, light-curve width, and color \citep{1993ApJ...413L.105P,1998A&A...331..815T}, which reduce the intrinsic dispersion in absolute peak magnitude to $\sim$0.1 mag, corresponding to a distance precision of $\sim$5\% per event \citep{2018ApJ...859..101S}. This precision, accumulated over large samples, has enabled the construction of Hubble diagrams extending to redshifts $z > 1$ and the measurement of the dark energy density and equation-of-state parameter $w$ to a few percent \citep{2014A&A...568A..22B,2024ApJ...973L..14D}. As cosmological surveys enter a new era of statistical power with the Vera Rubin Observatory's Legacy Survey of Space and Time (LSST; \citealt{2019ApJ...873..111I}), the Nancy Grace Roman Space Telescope \citep{2021arXiv211103081R}, and the Lazuli Space Observatory \citep{2026arXiv260102556R}, the discovery rate of SNe Ia will increase by orders of magnitude. In this regime, systematic uncertainties in SN\,Ia standardization rather than statistical ones will set the ultimate limits on the precision of cosmological inference, and their characterisation and mitigation becomes the central challenge of the field. At the same time, JWST is extending SN~Ia discoveries to much higher redshift, into epochs where the Universe is increasingly matter-dominated (e.g., \citealt{2025ApJ...979..250D,2025ApJ...981L...9P}). The progenitor environments in this regime are expected to differ from those of most local SNe~Ia, in particular through lower metallicities, younger stellar populations, and higher specific star-formation rates. These shifts may change the mapping between local environment, light-curve parameters, and standardized luminosity, turning environmental systematics into a redshift-dependent challenge for cosmology rather than a purely local effect.

A key source of such systematic uncertainty lies in the diversity of SN\,Ia progenitor systems and their parent stellar populations. Despite decades of effort, the physical origin of SNe Ia remains unresolved, including the nature of the progenitor system (e.g. the donor-star identity) and the explosion pathway (e.g. mass transfer and ignition conditions in the white dwarf; \citealt{2014ARA&A..52..107M,2023RAA....23h2001L}). 
\revdel{Two broad evolutionary mechanisms are commonly discussed: the single-degenerate (SD) path, in which a carbon-oxygen white dwarf accretes from a non-degenerate companion, and the double-degenerate (DD) path, in which two white dwarfs interact or merge after orbital decay (Whelan \& Iben 1973;Whelan \&Iben1984;Whelan \& Iben 1984). Recent work increasingly frames the problem in terms of the mass of the exploding white dwarf (near-Chandrasekhar versus sub-Chandrasekhar), rather than SD versus DD alone, because both evolutionary mechanisms may populate both mass regimes (see Ruiter \& Seitenzahl 2025). In this picture, the exploding WD mass is physically central because it sets the density structure and nucleosynthetic outcome: near-$M_{\rm Ch}$ explosions reach higher central densities, whereas sub-$M_{\rm Ch}$ explosions generally require an external trigger such as helium-shell ignition in double-detonation models (Livne1990; Livne 2010). These scenarios predict different delay-time distributions (DTDs; the time between progenitor formation and SN\,Ia explosion), especially in turn-on timescale and late-time slope (Mennekens et al. 2010; Mennekens et al. 2011). Measuring the DTD empirically, and in particular constraining the minimum delay time, therefore provides one of the most direct observational constraints on which explosion pathways dominate and over which progenitor mass ranges they operate.}
\revadd{The proposed channels, single- versus double-degenerate progenitor systems, and near- versus sub-Chandrasekhar explosions \citep{1973ApJ...186.1007W,1984ApJS...54..335I,1984ApJ...277..355W,2025A&ARv..33....1R}, predict different delay-time distributions (DTDs; the time between progenitor formation and explosion), particularly in their turn-on timescale and late-time slope \citep{2010A&A...515A..89M,2011MNRAS.417..408R}. Measuring the DTD empirically, and especially constraining the minimum delay time, therefore provides one of the most direct observational constraints on which explosion pathways dominate.}

Constraining the DTD through direct progenitor detection in pre-explosion imaging has proven largely unviable. The SN\,Ia rate in galaxies with existing deep, high-resolution imaging is low, and the expected luminosities of viable progenitor systems are far below those of the massive star progenitors of core-collapse SNe, which have been successfully identified in archival Hubble Space Telescope (HST) images in numerous cases \citep{2009ARA&A..47...63S,2017RSPTA.37560277V}. Pre-explosion constraints for SNe~Ia exist for only a handful of events and generally amount to luminosity upper limits rather than secure progenitor detections \citep{2011MNRAS.412.1441L,2014ApJ...790....3K,2014Natur.512...54M}. Outside the optical, direct progenitor detections for normal SNeIa remain rare and ambiguous e.g. SN~2007on, \citealt{2008Natur.451..802V,2008MNRAS.391..290R}), while deep radio, X-ray, and UV observations of nearby events have largely yielded upper limits that disfavour luminous supersoft and red-giant donor configurations rather than unambiguous detections \citep{2012ApJ...753...22B,2012ApJ...746...21H}. The path to large-scale DTD constraints must therefore run through indirect, statistical methods.

Environment and host galaxy studies have emerged as the most productive alternative route. The foundational insight is that the coeval stellar population surrounding a SN explosion site \revdel{encodes information about the progenitor age} \revadd{statistically constrains the progenitor-age distribution}, metallicity, and mass. By characterising this local population statistically across large SN samples, one can \revdel{reconstruct} \revadd{constrain} the progenitor properties \revadd{in an ensemble sense }without detecting any individual system. Early studies demonstrated that SN\,Ia rates per unit stellar mass are elevated in star-forming relative to passive galaxies \citep{2005A&A...433..807M,2006ApJ...648..868S}, consistent with a bimodal DTD with both prompt ($\lesssim 500$ Myr) and delayed ($\gtrsim 1$ Gyr) components. Subsequent work using the integrated stellar masses and star formation rates of host galaxies established that SNe Ia in more massive, passive hosts are systematically brighter after light-curve corrections (the so-called mass-step), a $\sim$0.06 mag offset that has now been confirmed across multiple independent datasets and analysis pipelines (e.g. \citealt{2010MNRAS.406..782S,2013ApJ...770..108C,2014A&A...568A..22B,2018ApJ...859..101S,2020MNRAS.494.4426S,2021MNRAS.501.4861K,2024ApJ...970...72U}). The physical interpretation of this step remains actively debated, with stellar age, metallicity, and dust all proposed as the underlying driver \citep{2013ApJ...764..191H,2013A&A...560A..66R,2021ApJ...909...26B}, but the mere existence of the step implies that the progenitor environment imprints itself measurably on the explosion, through a mechanism not fully captured by the standard light-curve width and color corrections.

Progress toward resolving this debate has been impeded by the use of integrated host galaxy properties as proxies for the local progenitor environment. A galaxy's total stellar mass or global star formation rate is an average over scales of tens of kpc and stellar populations spanning billions of years, and is a notoriously noisy tracer of the conditions at the specific location where a SN exploded. The advent of wide-field integral-field spectroscopy (IFS) has enabled a transition from integrated to spatially resolved host galaxy studies, isolating the stellar population at the SN site rather than averaging over the whole galaxy (e.g., \citealt{2014A&A...572A..38G,2016MNRAS.455.4087G,2018MNRAS.473.1359L,2018A&A...613A..35K}). These studies use emission-line ratios to derive gas-phase metallicities and star formation rates, and spectral synthesis of the stellar continuum to infer local stellar ages and star formation histories. This approach has yielded significant results: \cite{2020A&A...644A.176R} showed, using the local specific star formation rate measured within a projected 1 kpc radius of the SN position from H$\alpha$ emission in IFS data, that SNe Ia in locally star-forming environments are systematically fainter after standardization than those in locally passive environments, at a significance exceeding $5\sigma$. This local environmental step is larger than the global mass step, suggesting that the relevant physical driver operates on small spatial scales and is diluted when measured in integrated host properties. Similar conclusions have been reached independently using photometric proxies for local star formation in ground-based imaging \citep{2018A&A...615A..68R,2021A&A...649A..74N}.

A fundamental limitation persists across all these analyses: optical data alone is poorly suited to characterize the stellar populations most relevant to SN~Ia progenitor studies. In the rest-frame optical, the integrated light is dominated by evolved populations, while stars with ages of tens to a few hundreds of Myr contribute relatively little continuum flux and are difficult to disentangle from older components. This introduces severe degeneracies in stellar-population synthesis: at ages of $\sim$10--500 Myr and sub- to super-solar metallicities, different combinations of age, reddening, and metallicity produce nearly identical optical SEDs \citep{2003MNRAS.342..259D,2013ARA&A..51..393C,2016MNRAS.458..184L}. In the context of SN\,Ia host galaxies, varying the UV flux while keeping the optical spectrum fixed can shift best-fit ages by nearly an order of magnitude, implying large systematics in inferred DTD constraints. Crucially, UV data constrain the intermediate-age ($\sim$30--300 Myr) component that is weakly constrained by optical data, whereas H$\alpha$ mainly traces $\lesssim 10$ Myr star formation. This intermediate window is precisely where SN~Ia progenitor-delay constraints are expected to be most informative. At 2000--3000 \AA, hot young stars outshine old populations by orders of magnitude, and the UV flux evolves steeply with age (e.g., single stellar population models at 3 and 300 Myr differ in F275W by more than an order of magnitude), providing age leverage absent in optical-only fitting. We note, however, that binary products such as stripped stars can also contribute non-negligible UV flux at intermediate-to-old ages (e.g., \citealt{2019A&A...629A.134G,2022ARA&A..60..455E}). UV data therefore act as a key anchor for stellar-population synthesis, breaking the age--metallicity--reddening degeneracy and enabling more reliable recovery of the young stellar fraction at SN explosion sites \citep{2003MNRAS.342..259D,2016MNRAS.458..184L,2016MNRAS.457.4296W}.

The importance of UV data as a remedy for this degeneracy has already been recognised in the SN\,Ia context. \cite{2015ApJ...802...20R} used GALEX far-UV imaging for 77 SNe~Ia from the Constitution sample to classify local environments within a 2 kpc radius aperture, independently confirming the Hubble residual step found from H$\alpha$ IFS measurements \citep{2013A&A...560A..66R}, and demonstrating that UV-bright local environments identify the same population of prompt SNe Ia. However, GALEX carries two fundamental limitations that the current project is designed to overcome. First, its angular resolution of $\sim$5 arcsec corresponds to physical scales of $\sim$3--8 kpc at the redshifts of typical nearby SN\,Ia hosts, making it impossible to isolate the 1 kpc environment around the explosion site from the surrounding galaxy disc, the scale at which environmental effects are expected to be strongest \citep{2020A&A...644A.176R}. Second, GALEX's sensitivity is insufficient to detect faint or quiescent local environments: upper limits in UV-passive regions carry little constraining power, leaving the optically inferred ages of locally passive hosts essentially unconstrained. UV coverage at the angular resolution and sensitivity required to work at sub-kpc scales in nearby galaxies is uniquely provided by HST's WFC3/UVIS camera, which reaches point-source depths of \revdel{$M_{\rm F275W}$} \revadd{$m_{\rm F275W}$}$\,\sim 25.8$ AB mag in a single orbit and delivers a FWHM of $\sim$0.07 arcsec, resolving physical scales of $\lesssim 0.1$ kpc at $z \sim 0.02$. Critically, the HST upper limits in UV-passive environments are deep enough to genuinely suppress the young stellar component in spectral-synthesis fitting: non-detections provide physically meaningful constraints on recent star formation that are simply unavailable from GALEX data.

In this paper we present the analysis of HST Cycles 29 \& 30 SNAP proposals (IDs 16741 \& 17179; PI: Galbany), which obtained deep WFC3/UVIS F275W imaging of 340 SN\,Ia host galaxies, each with pre-existing or planned VLT/MUSE IFS, constructing the largest and most homogeneous UV+IFS dataset of SN\,Ia environments assembled to date. For each galaxy, we perform consistent photometric and spectroscopic measurements within a 1 kpc radius aperture centred on the SN position, and fit the UV photometry and optical spectrum jointly with stellar population synthesis models to derive local stellar ages, masses, star formation rates, gas-phase metallicities, and dust attenuation. For the subset of SNe Ia with well-sampled light curves, standardised distances and Hubble residuals enable a direct connection between local environment and SN\,Ia cosmological performance. This combination of resolved UV imaging, IFS-derived stellar population parameters, emission-line diagnostics, and SN light-curve properties at a common physical scale of 1 kpc radius \revdel{constitutes a qualitative advance over all previous SN\,Ia host environment studies} \revadd{represents a substantial and complementary advance over previous SN\,Ia host environment studies}, and provides a unique dataset to simultaneously constrain SN\,Ia progenitor physics and quantify the environmental systematics that will limit next-generation cosmological analyses. Throughout we adopt a flat $\Lambda$CDM cosmology \citep{Planck2020Cosmo} implemented in \texttt{astropy}, with $H_0 = 70$ km s$^{-1}$ Mpc$^{-1}$.

%%%%%%%%%%%%%%%%%%%%%%%%%%%%%%%%%%%%%%%%%%%%%%%%%%%%%%%
%%%%%%%%%%%%%%%%%%%%%%%%%%%%%%%%%%%%%%%%%%%%%%%%%%%%%%%
%%%%%%%%%%%%%%%%%%%%%%%%%%%%%%%%%%%%%%%%%%%%%%%%%%%%%%%

\section{Data sample}\label{sample}

\subsection{HST ultraviolet imaging}

Ultraviolet imaging was obtained from two HST/WFC3 UVIS F275W-filter SNAP programmes designed to build a homogeneous UV imaging sample of nearby SN~Ia host galaxies with existing or planned integral-field spectroscopy from the All-weather MUse Supernova Integral field Nearby Galaxies survey (AMUSING; \citealt{2016MNRAS.455.4087G,2020AJ....159..167L}) using the MUSE instrument at the 8.2m Very Large Telescope (VLT): 156 images come from programme 16741\footnote{\href{https://archive.stsci.edu/proposal_search.php?id=16741&mission=hst}{\it A public UV snapshot survey of SN\,Ia hosts in IFS data.}} and 190 from 17179\footnote{\href{https://archive.stsci.edu/proposal_search.php?id=17179&mission=hst}{\it A public HST-UV snapshot survey of SN\,Ia host galaxies with pre-existing optical IFS}}, both led by our team. Targets were selected by known SN\,Ia events rather than from a pre-defined host-galaxy catalog, resulting in a broad distribution of host properties such as morphology, stellar mass, and metallicity. At the same time, this sample remains shaped by the discovery and spectroscopic follow-up selection functions of the contributing transient surveys. Relative to a strictly galaxy-targeted host pre-selection, however, the SN-triggered strategy reduces direct host-property cuts and is therefore well suited to testing whether SN\,Ia properties correlate with the local environment at the explosion site. These two programmes are twin proposals to 16287, which was designed using the same general strategy but targeted the host galaxies of core-collapse supernovae (CCSNe) rather than those of SNe Ia. \cite{2025A&A...693A..39I} presented the corresponding HST UV sample for CCSN environments and described the parallel sample construction and image-analysis framework used in that context.

The observing strategy used three dithered UVIS exposures per visit, balancing depth and image quality. Exposure times were adjusted according to host angular extent and aperture configuration, with typical integrations of $3\times400$ to $3\times500$~s. Depth estimates from the Exposure Time Calculater (ETC) indicate that these integrations with this configuration are sufficient to detect faint recent star-forming structures in nearby hosts and to provide useful constraints for stellar-population modeling when combined with optical IFS. The observations were obtained with low post-flash (\texttt{FLASHLVL}=20) to mitigate charge-transfer-efficiency (CTE) losses in low-background UV imaging. For all targets we use the final reduced \texttt{\_drc} products retrieved from the HST archive, that is, the combined drizzled images corrected for geometric distortion and detector charge-transfer effects. In contrast to \cite{2025A&A...693A..39I}, we do not apply an additional astrometric recalibration to the HST images. In that work the main goal was to assess emission at the precise SN position, which requires a more stringent astrometric treatment. Here, however, our measurements are based on matched 1~kpc radius apertures centred on the SN coordinates, and the native astrometry of WFC3/UVIS \texttt{\_drc} products is adequate for that purpose (conservatively $\lesssim0.2''$, and often better for Gaia-calibrated products, corresponding to $\sim0.08$--$0.12$ kpc over $z\sim0.02$--0.03). Checks of the smallest-aperture, highest-redshift cases do not indicate systematic HST-optical misalignments large enough to affect our aperture-based measurements.

\begin{figure}[t]
\centering
\includegraphics[width=\columnwidth]{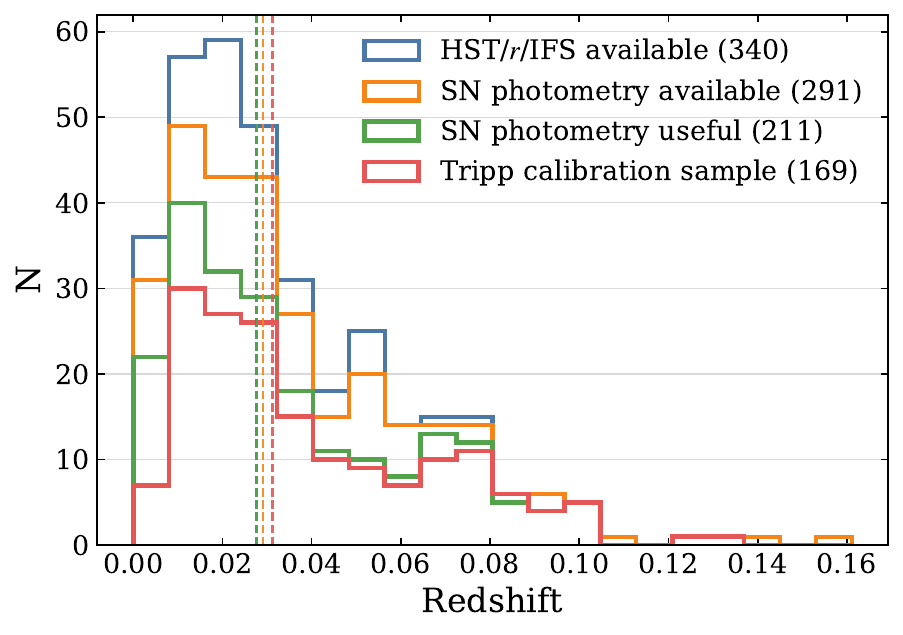}
\caption{Redshift distribution of the SN host galaxy samples used in this work. The blue histogram shows the initial sample with HST UV imaging and IFS available, the orange histogram shows the subset with available SN~Ia light curves, the green histogram shows the final sample of SNe Ia with useful light-curves for template fitting, and the red histogram the calibration sample used to determine the Tripp standardization parameters. Dashed vertical lines mark the median redshift of each subsample.}
\label{fig:redshift_distribution}
\end{figure}

From the 346 SNAP-programmes images, 31 targets, all from programme 17179, are not included in the present analysis because the corresponding MUSE observations are still planned but have not yet been executed. An additional search of the HST archive by coordinates of all galaxies in the AMUSING sample provided F275W imaging for 23 more SN locations. The resulting parent compilation therefore comprises 338 HST/WFC3 F275W drizzled images covering 340 SN positions. The redshift distribution of this compilation spans $z=0.0018$ to $z=0.1564$, with a median redshift of $z=0.0276$, as shown in Fig.\,~\ref{fig:redshift_distribution}.

\subsection{Optical $r$-band imaging}\label{data:optical}

Optical imaging at the SN position was obtained with \texttt{hostphot}\footnote{\href{https://hostphot.readthedocs.io/en/latest/}{https://hostphot.readthedocs.io/en/latest/}} (v. 3.1.4; \citealt{2022JOSS....7.4508M}). For each target, \texttt{hostphot} queries the Legacy Survey \citep{2019AJ....157..168D}, the Dark Energy Survey \citep{2021ApJS..255...20A}, Pan-STARRS \citep{2016arXiv161205560C}, and SkyMapper \citep{2024PASA...41...61O}, and checks for available $r$-band imaging. All 340 SN positions have $r$-band imaging available from at least one of these surveys. Because many SN sites are covered by more than one optical survey, these data also provide a useful cross-check on the consistency of the local optical photometry. The corresponding survey coverage is 339 objects in Legacy Survey, 97 in DES, 240 in Pan-STARRS, and 309 in SkyMapper. In addition, 62 objects have usable imaging in all four surveys, 188 in three surveys, 83 in two surveys, and 7 in only one survey. 

\subsection{MUSE integral-field spectroscopy}

Optical integral-field spectroscopy for the host galaxies in our sample was obtained from the AMUSING survey, a long-term VLT/MUSE programme designed to build a large and statistically useful sample of spatially resolved spectroscopy of nearby SN host galaxies, by taking advantage of suboptimal observing conditions at Cerro Paranal observatory. These MUSE IFS observations provide the spectroscopic counterpart to the HST UV imaging described above and enable the characterization of the local ionized-gas and stellar-population properties at the SN locations on matched physical scales.

MUSE delivers a contiguous optical datacube for each target host galaxy, providing a spectrum at every spatial element of $0\farcs2 \times 0\farcs2$ across a field of view of approximately $1\arcmin \times 1\arcmin$. This field is sufficient in most cases to cover the central regions of the host galaxy together with the SN environment, while the spectral coverage includes the main optical stellar-continuum and nebular-emission features required for local environmental studies. In particular, the MUSE wavelength range includes the principal Balmer lines and the strongest forbidden transitions used to estimate extinction, star-formation activity, ionization conditions, and gas-phase metallicity as inferred from a region of 1 kpc-radius centered on the SN position.

A corresponding MUSE datacube is available for all 338 HST/WFC3 F275W images, covering all 340 SN positions. These cubes define the parent spectroscopic dataset used for the local-environment analysis presented in Sect.~\ref{sec:anamuse}.

\subsection{SN\,Ia light curves}\label{sec:lcdata}

For the 340 SNe Ia in the matched HST+IFS compilation, we carried out a systematic search for available SN light curves. By construction, most of the light curves in our compiled sample come from the Carnegie Supernova Project (CSP), including both the initial CSP-I campaign (2004--2009; \citealt{2017AJ....154..211K}) and its CSP-II extension (2011--2015; \citealt{2019PASP..131a4001P}). In addition, we carried out a thorough search in the literature for publicly available light-curves for the remaining sample. For a small number of more recent objects with incomplete or absent compiled light curves, we also queried the Zwicky Transient Facility (ZTF\footnote{\href{https://ztfweb.ipac.caltech.edu/cgi-bin/requestForcedPhotometry.cgi}{https://ztfweb.ipac.caltech.edu/cgi-bin/requestForcedPhotometry.cgi}}; \citealt{2019PASP..131a8002B}) and the Asteroid Terrestrial-impact Last Alert System (ATLAS\footnote{\href{https://fallingstar-data.com/forcedphot/}{https://fallingstar-data.com/forcedphot/}}; \citealt{2018PASP..130f4505T}) force-photometry services \citep{2019PASP..131a8003M,2021TNSAN...7....1S} at the SN coordinates. The recovered force-photometry files were baseline-corrected, filtered by a minimum signal-to-noise (S/N) threshold of three, and converted into a common \texttt{sncosmo} \citep{Barbary2016sncosmo} format in preparation for the fitting analysis described in Sect.~\ref{sec:anasn}. 

This search yielded 291 SNe with at least one photometric point in at least one photometric band, while 49 objects have no public photometry available in any form. These 291 SNe are covered by 290 unique HST/F275W images, with one HST pointing containing two SN sites. Their redshift distribution is shown in Fig.\,~\ref{fig:redshift_distribution}, with a median redshift of $z=0.0291$. Of these 291 SNe with available photometry, 217 are from the CSP, comprising 89 objects from CSP-I and 128 from CSP-II. Three objects were supplemented with usable force-photometry products, two from ZTF and one from ATLAS, while the remaining 71 were assembled from a heterogeneous set of public low-redshift SN data in the literature.

\revdel{We then carried out a visual quality-control inspection of the assembled light curves to identify cases where public photometry exists but the phase coverage is inadequate for template fitting (e.g. only data on the rise or only at late epochs). At this stage we exclude objects for which the available measurements are too sparse in time, too poorly sampled around maximum light, too limited in filter coverage, or otherwise dominated by late-time or secondary-maximum data in a way that prevents a reliable fit to the primary SN~Ia light-curve peak. After this visual check, 80 objects are discarded and 211 SNe remain for the subsequent fitting analysis.} \revadd{We applied light-curve quality requirements to ensure that the SALT3-NIR shape and color parameters are well constrained. A light curve is retained only if it (i) includes at least two photometric bands; (ii) has its first epoch no later than the time of maximum, so that the peak is constrained rather than extrapolated; and (iii) includes at least one post-maximum epoch sampling the decline. Individual force-photometry epochs are additionally required to have $\mathrm{S/N}>3$. For the 211 retained light curves these requirements correspond to a median of nine photometric bands (minimum two) and 132 epochs (minimum 20), a first epoch at a median rest-frame phase of $-6$~d, and coverage extending to a median of $+52$~d ($95\%$ reach at least $+15$~d). Objects lacking near-maximum coverage, sampling only the late-time tail or the secondary maximum, or with too few epochs to constrain $(x_0,x_1,c)$ fail these requirements. After applying them, 80 objects are discarded and 211 SNe remain for the subsequent fitting analysis.} 
In this case, all retained SNe have a unique HST/F275W image. This final sample is still dominated by CSP data, with 180 of the 211 useful light curves, including 69 CSP-I and 111 CSP-II SNe Ia, while the remaining 31 come from the literature. The median redshift of the final curated light-curve sample is $z=0.0277$, as shown in Fig.\,\ref{fig:redshift_distribution}. Table~\ref{tab:sample_flow} summarizes the main sample-construction stages used in this analysis.

\begin{table}[t]
\caption{Summary of the sample-construction stages used in this work.}
\label{tab:sample_flow}
\centering
\resizebox{\columnwidth}{!}{%
\begin{tabular}{lcc}
\hline\hline
Sample stage & Galaxies & SN locations \\
\hline
HST/F275W drizzled images                    & 338       & 340 \\
$r$-band optical imaging                     & 340       & 340 \\
MUSE IFS datacubes                           & 338       & 340 \\
SN available light curves                    & 290       & 291 \\
\hline
Successful HST/F275W photometry              & 335       & 337 \\
Successful $r$-band optical photometry       & 338       & 340 \\
Successful MUSE 1~kpc spectrum               & 335       & 337 \\
Successful STARLIGHT SSP fit                 & 333       & 335 \\
Good \revdel{SNooPy} \revadd{\texttt{sncosmo}}/SALT3-NIR light-curve fits       & 211       & 211 \\
Pass cosmology light-curve cuts              & 169       & 169 \\
\hline
\end{tabular}}
\end{table}

%%%%%%%%%%%%%%%%%%%%%%%%%%%%%%%%%%%%%%%%%%%%%%%%%%%%%%%
%%%%%%%%%%%%%%%%%%%%%%%%%%%%%%%%%%%%%%%%%%%%%%%%%%%%%%%
%%%%%%%%%%%%%%%%%%%%%%%%%%%%%%%%%%%%%%%%%%%%%%%%%%%%%%%

\begin{figure*}[!t]
\centering
\includegraphics[width=\textwidth]{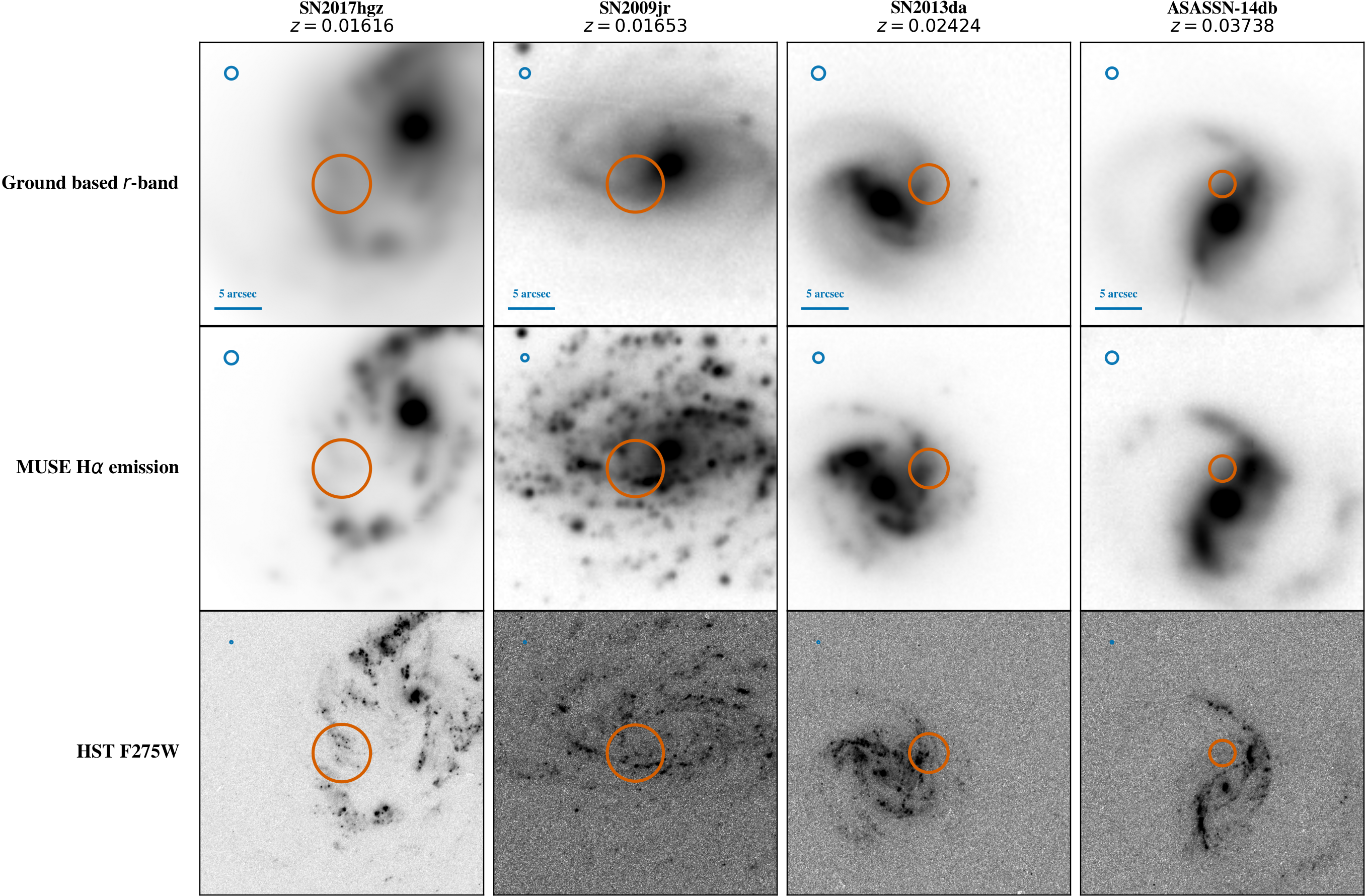}
\caption{Representative environmental-imaging examples ordered left to right by increasing SN redshift: SN~2017hgz, SN~2009jr, SN~2013da, and ASASSN-14db. Each column corresponds to one SN, while the rows show, from top to bottom, the adopted optical $r$-band cutout, the observed-frame MUSE H$\alpha$ map collapsed over $\pm10$\AA\ around the redshifted line\revadd{ (a narrow-band map containing H$\alpha$ plus the underlying continuum)}, and the HST/F275W cutout. All panels are reprojected onto the same north-up $30\arcsec \times 30\arcsec$ field centered on the SN position, and the orange circle marks the common 1kpc-radius aperture used throughout the analysis. The blue circle in the upper-left corner of each panel indicates the image PSF FWHM for that specific dataset, highlighting the resolution differences across bands. These examples illustrate the analysis products and aperture-matching strategy, rather than the host-morphology distribution of the full sample (shown in Appendix \ref{app:figs}).}
\label{fig:main_grid}
\end{figure*}

\section{Analysis}

\subsection{UV and optical photometry}\label{sec:uvopt}

We measure the local UV and optical photometry at each SN position using the same circular physical 1~kpc-radius aperture adopted for the MUSE spectral extraction. The use of a common 1~kpc-radius aperture ensures that the UV and optical fluxes are measured on matched spatial scales and can therefore be interpreted together as tracers of the same local environment. We adopt 1 kpc as a practical compromise between spatial resolution, S/N, and homogeneous measurements across the sample. Smaller apertures are resolution-limited, while larger apertures increasingly dilute local environmental contrasts. The UV measurements are obtained from the HST/WFC3 UVIS F275W images, while the optical measurements are derived from the adopted $r$-band images following the survey-priority scheme: Legacy Survey, DES, Pan-STARRS, and SkyMapper. The adopted $r$-band measurement is the first one in the above-listed sequence with $\mathrm{S/N}\geq3$ in the 1~kpc-radius aperture. If no survey reaches that threshold, we adopt the first available 3$\sigma$ upper limit in the same priority order. While all 340 SN environments end up having an $r$-band detection (330) or an upper limit (10), we are able to obtain UV detections in 235 cases and 102 upper-limits. Three locations have no usable UV measurement: SN~2005cf falls outside of the footprint in one of the two HST images available, while SN~2012Z and SN~2007af were manually rejected after visual inspection because the projected 1~kpc-radius aperture extends beyond the edge of the F275W image (see Appendix \ref{app:figs}). 

\subsection{MUSE H$\alpha$ 2D maps}

The MUSE datacubes are used in two different ways. First, we construct two-dimensional H$\alpha$ maps from the datacubes in order to visualize the spatial distribution of the ionized gas around the SN site. To construct the H$\alpha$ maps, we use the SN redshift to determine the expected observed-frame wavelength of H$\alpha$ and collapse the datacube over a narrow interval of $\pm10$~\AA\ around that wavelength to avoid including [N\,{\sc ii}] $\lambda\lambda$6548,84 contamination. This produces a \revdel{two-dimensional map of the local H$\alpha$ surface-brightness distribution} \revadd{narrow-band map containing H$\alpha$ emission plus the underlying local continuum} that can be directly compared with the HST/F275W and optical $r$-band imaging.
These maps are used primarily as a visualization and quality assessment, allowing us to examine how the local UV emission, optical continuum, and nebular line emission are distributed around the SN environment. Figure~\ref{fig:main_grid} shows four representative systems ordered by SN redshift. For each object we display the same north-up $30\arcsec$ cutout geometry of the adopted optical $r$-band image, the observed-frame MUSE H$\alpha$ map, and the HST/F275W image. All panels are shown with the same 1~kpc-radius aperture, emphasizing the matched-aperture strategy used throughout the local-environment analysis. 

\begin{figure*}[!t]
\centering
\includegraphics[width=\textwidth]{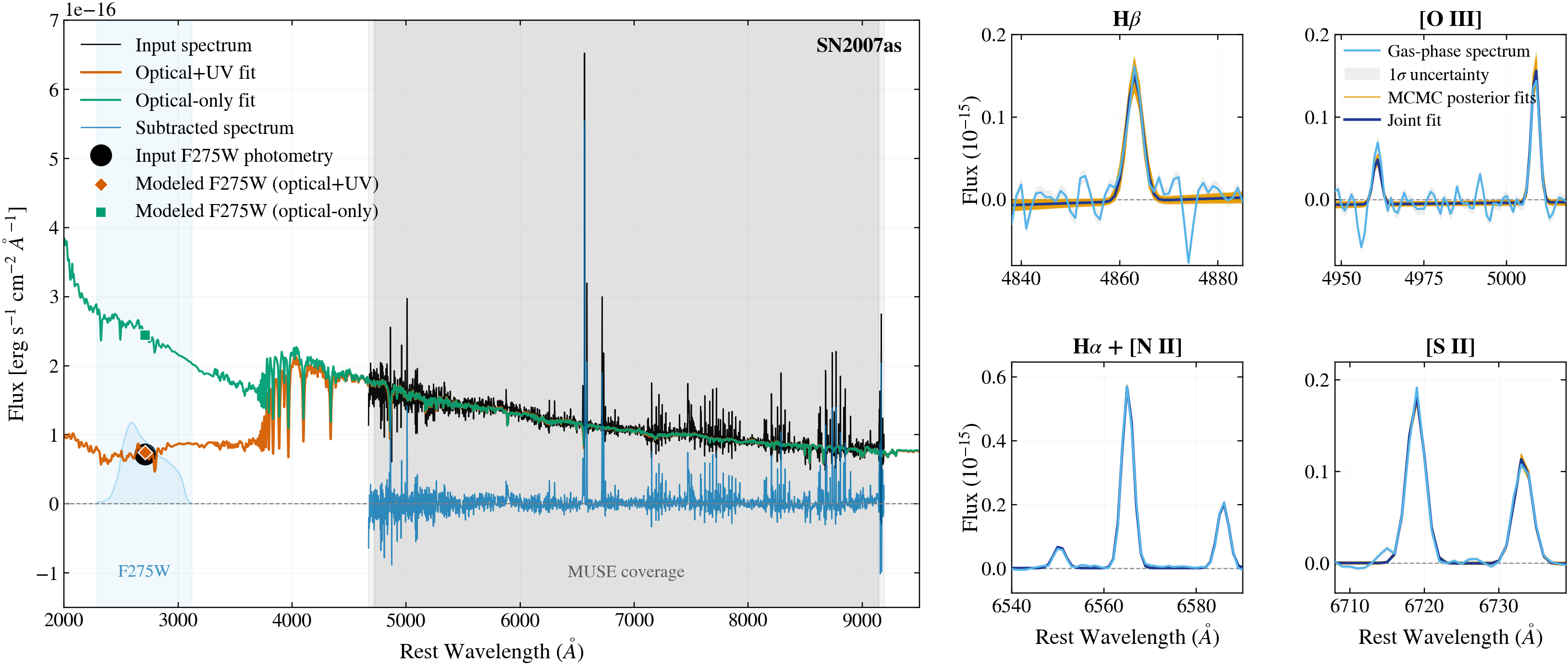}
\caption{Example of the 1~kpc MUSE spectral analysis at the SN position for SN~2007as. The large panel on the left compares the optical-only and joint optical+F275W \texttt{STARLIGHT} fits to the observed rest-frame spectrum, together with the continuum-subtracted residual spectrum and the observed and modeled F275W photometric constraint. Residual narrow telluric features are interpolated in the rest-frame spectrum for visual clarity. The four panels on the right show the local Bayesian fits to H$\beta$, [O\,{\sc iii}], H$\alpha$+[N\,{\sc ii}], and [S\,{\sc ii}], including the continuum-subtracted gas-phase spectrum, the posterior MCMC realizations, and the median joint fit.}
\label{fig:muse_spec_example}
\end{figure*}

\subsection{MUSE 1~kpc environment spectra}\label{sec:anamuse}

Beyond these two-dimensional maps, we extract a one-dimensional MUSE spectrum at each SN position using the same circular aperture with a physical radius of 1~kpc used to measure photometry. For each SN, we compute the corresponding angular size at the host-galaxy redshift, project that aperture onto the MUSE datacube through the cube WCS, and sum the flux within it at every wavelength slice. The associated variance spectrum is propagated from the cube \texttt{STAT} extension using the exact fractional aperture weights. Three objects are excluded from the MUSE environment-analysis sample, two because the projected 1~kpc-radius aperture falls outside the usable MUSE footprint (SN~2005cf and SN~2017cbv) and one because the extracted local spectrum is contaminated by SN light (SN~2020xyw).

\subsection{Stellar population parameters}\label{sec:sp}

The extracted spectra are prepared for stellar-population fitting by correcting for Milky Way foreground reddening at the SN coordinates using the SFD dust maps \citep{Schlafly2011} and a \citet{Fitzpatrick1999} reddening law with $R_V=3.1$, shifting to the rest frame, and resampling onto a uniform 1~\AA\ grid. In addition to the optical spectrum, we include the local HST/WFC3 F275W photometry measured within the same projected 1~kpc-radius aperture as an additional constraint  on the stellar-population fit. We then fit these data with \texttt{STARLIGHT} (\texttt{v06r02}; \citealt{2010MNRAS.403.1036C}) in a joint spectro-photometric mode \citep{2016MNRAS.458..184L}. The code models the data as a linear combination of simple stellar populations of different ages and metallicities, while simultaneously solving for dust attenuation and stellar kinematics. For the \texttt{STARLIGHT} fits we adopt the 45-element \citet{BruzualCharlot2003} single stellar-population models, spanning 15 ages\revadd{ logarithmically spaced from 1~Myr to 13~Gyr ($\log t/\mathrm{yr}=6.0$--$10.1$)} and three metallicities ($Z = 0.004$, 0.02, and 0.05; $\approx0.2, 1,$ and $2.5\,Z_\odot$), and built with a \cite{2003PASP..115..763C} initial mass function (IMF). For the stellar-continuum analysis we fit each rest-frame spectrum over its available wavelength range after trimming 50~\AA\ from both the blue and red edges, adopting a \citet{Fitzpatrick1999} extinction law for the attenuation component, and masking strong sky-line regions. The HST F275W measurement is incorporated using the observed-frame F275W filter transmission curve and the source redshift to compare the model spectral energy distribution (SED) with the measured UV magnitude. 
\revadd{For UV non-detections, the F275W constraint is passed to \texttt{STARLIGHT} as a one-sided upper limit; the spectrophotometric \texttt{STARLIGHT} implementation natively supports such one- and two-sided photometric constraints \citep{2016MNRAS.458..184L}. The $3\sigma$ limiting magnitude is supplied and flagged as an upper limit in the photometric-constraint block, so that the fit is penalized only when the model F275W flux exceeds the limit and incurs no penalty when it falls below it.} % The limit is therefore not treated as a zero-flux measurement or as a symmetric $3\sigma$ flux, but acts to suppress, rather than to fix, the young stellar component.}

From each successful fit we recover the main stellar-population parameters, including the current and total formed stellar mass, the light- and mass-weighted ages and metallicities, the best-fitting stellar attenuation, and the stellar kinematics. To estimate statistical uncertainties on the \texttt{STARLIGHT}-derived parameters, we generated 50 Monte Carlo realizations of each input spectrum by perturbing the flux in every wavelength with Gaussian noise drawn from the flux uncertainty at that wavelength. Each realization was refit with \texttt{STARLIGHT}, and the uncertainties were taken from the distribution of recovered values, using the 16th and 84th percentiles to define the lower and upper errors.

For comparison, we also perform an optical-only \texttt{STARLIGHT} fit, which allows us to evaluate the impact of adding the UV photometry on the recovered stellar-population properties. Figure~\ref{fig:muse_spec_example} shows an example of these two fits for the SN~2007as local environment. Both fits reproduce the optical spectrum well, but they diverge strongly in the UV, where the optical-only fit overpredicts the F275W flux. This leads to substantially different luminosity-weighted ages: the optical-only fit gives $\log t_{L,\mathrm{opt}} = 8.00 \pm 0.11$ dex ($1.00\times10^8$ yr), while the joint fit gives $\log t_{L,\mathrm{UV+opt}} = 8.58 \pm 0.03$ dex ($3.81\times10^8$ yr). 

In the following analysis we adopt the joint UV+optical fit as the baseline. This procedure provided 335 successful fits after excluding the three objects without usable UV measurements (SN~2005cf, SN~2007af, and SN~2012Z) and the three objects without a usable local 1~kpc MUSE spectrum for the joint analysis (SN~2005cf, SN~2017cbv, and SN~2020xyw).

\subsection{Gas phase parameters}\label{sec:gas}

The UV+optical spectrum best stellar model is then subtracted from the observed spectrum to obtain the gas-phase spectrum used for measuring line fluxes and perform the nebular analysis, and also a normalized spectrum to measure equivalent widths. Emission and absorption features are measured with a two-stage Gaussian-fitting scheme. We first obtain a deterministic fit in each spectral window around the feature to initialize the parameters and then run a Bayesian fit around that solution to recover posterior constraints on line flux, velocity, and width. For the principal blended groups we also impose physically motivated constraints, such as fixing the [O\,{\sc iii}] $\lambda5007/\lambda4959$ and [N\,{\sc ii}] $\lambda6583/\lambda6548$ doublets to their theoretical ratios (2.98 and 2.96, respectively; e.g. \citealt{2000MNRAS.312..813S,2006agna.book.....O}), while adopting bounded ratios between 1 and 2 for the Na\,{\sc i}\, doublet absorption pair (optically thick to thin limit; e.g. \citealt{1978ppim.book.....S,2011piim.book.....D}). Figure~\ref{fig:muse_spec_example} also shows the resulting continuum-subtracted gas-phase spectrum, and the Bayesian Gaussian fits to the principal emission-line complexes used in the subsequent nebular analysis.

After measuring the Balmer lines, we estimate the local nebular color excess from the observed H$\alpha$/H$\beta$ ratio following the standard nebular-extinction formalism \citep{Osterbrock2006,Calzetti2000},
\begin{equation}
E(B-V)_{\mathrm{gas}} =
\frac{2.5}{k(\mathrm{H}\beta)-k(\mathrm{H}\alpha)}
\log_{10}\left[
\frac{(\mathrm{H}\alpha/\mathrm{H}\beta)_{\mathrm{obs}}}{2.86}
\right],
\end{equation}
assuming case-B recombination, with an intrinsic ratio of $(\mathrm{H}\alpha/\mathrm{H}\beta)_0 = 2.86$, appropriate for $T_e = 10^4\,\mathrm{K}$ and $n_e \approx 100\,\mathrm{cm}^{-3}$. Here, $k(\mathrm{H}\beta)$ and $k(\mathrm{H}\alpha)$ are the wavelength-dependent reddening coefficients, $k(\lambda)\equiv A_\lambda/E(B-V)$, evaluated at H$\beta$ and H$\alpha$ from the adopted \citet{Fitzpatrick1999} extinction curve (with $R_V=3.1$). This is used to deredden all fitted emission-line fluxes using the same \citet{Fitzpatrick1999} extinction law adopted for the stellar-continuum analysis. The Balmer ratio is unavailable for seven objects because the local H$\alpha$ line is not reliably recovered, so that the Balmer decrement cannot be measured\revdel{ and the emission line fluxes remain uncorrected}. \revadd{These objects therefore have no H$\alpha$-based measurements and are excluded from all H$\alpha$-based diagnostics, but their stellar-population and UV-based quantities are unaffected.} We then compute the main local emission-line diagnostics from the extinction-corrected measurements, including emission line ratios, oxygen-abundances, and the local H$\alpha$-based star-formation rate (SFR) from the extinction-corrected H$\alpha$ luminosity using the \citet{1998ARA&A..36..189K} calibration,
\begin{equation}
\mathrm{SFR}_{\mathrm{H}\alpha}\;[M_\odot\,\mathrm{yr}^{-1}] = 5.5 \times 10^{-42}\, L_{\mathrm{H}\alpha}\;[\mathrm{erg}\,\mathrm{s}^{-1}],
\end{equation}
which assumes continuous star formation and includes a factor of $\sim 0.7$ to convert from a Salpeter to a \cite{2003PASP..115..763C} IMF \citep{KennicuttEvans2012}. From this we derive the corresponding H$\alpha$ SFR surface density, $\Sigma_{\mathrm{SFR,H}\alpha}$, by dividing the measured SFR by the physical area of the adopted aperture, and the H$\alpha$ specific SFR, ${\rm sSFR}_{\mathrm{H}\alpha}$, by dividing by the local stellar mass. In parallel, we also derive equivalent widths for the main emission and absorption features from the continuum-normalized spectra. Uncertainties on these derived quantities are propagated through Monte Carlo realizations of the measured line-flux posteriors, so that the reported errors account for both the line-measurement uncertainties and the uncertainty in the Balmer-decrement extinction correction.

\revadd{To flag environments whose ionized gas is not dominated by star formation, and for which the H$\alpha$ luminosity is therefore an unreliable SFR tracer, we classify each local spectrum on the [N\,{\sc ii}]-based BPT diagram \citep{1981PASP...93....5B} using the $\log(\mathrm{[O\,{\sc iii}]}\,\lambda5007/\mathrm{H}\beta)$ versus $\log(\mathrm{[N\,{\sc ii}]}\,\lambda6583/\mathrm{H}\alpha)$ line ratios measured in the same 1~kpc aperture. Environments above the \citet{2001ApJ...556..121K} maximum-starburst line are labelled AGN-like, those between the \citet{2003MNRAS.346.1055K} and \citet{2001ApJ...556..121K} lines composite, and those below the \citet{2003MNRAS.346.1055K} line star-forming. 
Because the spectra are extracted at the SN position rather than at the galaxy nucleus, this flags line ratios inconsistent with pure star formation, which may also arise from LINER-like emission powered by hot evolved stars or diffuse ionized gas, rather than necessarily indicating a bona-fide active nucleus.
An environment is classified when all four lines ([O\,{\sc iii}]\,$\lambda5007$, H$\beta$, [N\,{\sc ii}]\,$\lambda6583$ and H$\alpha$) yield a reliable, finite-flux Gaussian fit, that is, the two-stage fit returns a finite flux and posterior for each. The classification serves only to identify and remove environments where H$\alpha$ is an unreliable SFR tracer, so this flag is deliberately conservative. Of the 337 environments with a MUSE spectrum, 228 are star-forming, 63 composite and 39 AGN-like, while the remaining 7 lack a reliable H$\alpha$/[N\,{\sc ii}] fit and are unclassified. The three environments in the 340-object parent sample without a usable MUSE spectrum are likewise unclassified, giving 10 unclassified in total. Only the 39 AGN-like environments are excluded from the UV--H$\alpha$ SFR comparison, while composite environments are retained  \citep{2014A&A...563A..49S}. }

\subsection{Photometric parameters}

The HST/F275W and $r$-band fluxes measured within the common 1~kpc-radius aperture are first corrected for Milky Way foreground extinction using the recalibrated dust maps of \citet{Schlafly2011} together with the \citet{Fitzpatrick1999} extinction law. We then correct the photometry for host-galaxy attenuation using the stellar-continuum attenuation derived from the joint UV+optical \texttt{STARLIGHT} fit, parameterized by $A_V^\star$. For a given bandpass, the internal attenuation is computed as
\begin{equation}
A_\lambda = k_\lambda\,\frac{A_V^\star}{R_V},
\end{equation}
where $R_V=3.1$ and $k_\lambda \equiv A_\lambda/E(B-V)$ is evaluated from the adopted \citet{Fitzpatrick1999} extinction curve. In particular, we adopt $k_{\rm F275W}=6.253$ and $k_r=2.553$, which are then used to derive the host-corrected F275W$-r$ color and UV-based quantities. The corrected UV flux is available for all 335 environments, comprising 235 direct detections and 100 upper limits. The corresponding host-corrected F275W$-r$ color can be constrained for 330 environments, including 231 direct detections and 99 one-sided limits, while the remaining 5 cases have both UV and $r$ constrained only as limits and therefore do not yield a usable one-sided color constraint.

From the corrected UV flux we derive a set of local continuum-based environmental quantities for the common 1~kpc-radius aperture. We first convert the apparent magnitudes into luminosity-based quantities using the galaxy redshift and the adopted cosmology. Given the low-redshift nature of the sample, we do not force the measurements into a common template-dependent rest-frame passband through a conventional bandpass K-correction. Instead, we compute the emitted monochromatic luminosity density directly as
\begin{equation}
L_{\nu,\mathrm{rest}} =
\frac{4\pi D_L^2}{1+z}\,f_{\nu,\mathrm{corr}},
\end{equation}
where $D_L$ is the luminosity distance and $f_{\nu,\mathrm{corr}}$ is the Milky Way- and host-corrected observed flux density. This expression accounts for the leading redshift term required to transform the observed flux density into the emitted luminosity density \citep{Hogg2002}. In practice, the effective rest-frame wavelength is $\lambda_{\mathrm{rest}}=\lambda_{\mathrm{F275W}}/(1+z)$, which remains within the near-UV regime over the full redshift range of the sample. From this UV luminosity density we derive the UV luminosity surface density
\begin{equation}
\Sigma_{\mathrm{UV}} = \frac{L_{\nu,\mathrm{rest}}}{A},
\end{equation}
where $A=\pi(1~\mathrm{kpc})^2$ is the physical area of the adopted local aperture. Because all measurements are performed on the same physical scale, $\Sigma_{\mathrm{UV}}$ is simply a fixed rescaling of $L_{\nu,\mathrm{rest}}$, but it provides a more natural quantity when comparing local UV measurements on a per-unit-area basis.

We then convert the UV luminosity density into a star-formation rate using a \cite{2003PASP..115..763C} IMF UV calibration,
\begin{equation}
\mathrm{SFR}_{\mathrm{UV}}\;[M_\odot\,\mathrm{yr}^{-1}] =
1.08\times10^{-28}\,L_{\nu,\mathrm{rest}}\;[\mathrm{erg\,s^{-1}\,Hz^{-1}}],
\end{equation}
from \citet{2007ApJS..173..267S}. This calibration is appropriate for UV continuum measurements in the $\sim1500$--$2800$~\AA\ range and assumes continuous star formation over $\sim10^8$~yr timescales. Similarly, we derive the UV SFR surface density, $\Sigma_{\mathrm{SFR,UV}}$, by dividing the local UV-based SFR by the area of the adopted aperture, and the UV specific SFR, ${\rm sSFR}_{\mathrm{UV}}$, by dividing by the local stellar mass from the stellar-population fit.

The host-corrected F275W$-r$ color and the UV-derived $\log\,\Sigma_{\rm SFR,UV}$ measured within the same 1~kpc apertures complement the gas-phase diagnostics from the MUSE spectra by providing an independent continuum-based view of recent star formation and dust on matched local scales. In particular, the UV luminosity and UV-based SFR trace recent star formation over longer timescales than H$\alpha$, while the corrected F275W$-r$ color, $L_\nu$, and $\Sigma_{\mathrm{UV}}$ provide additional constraints on the recent stellar population and local dust content.

\begin{figure}[!t]
\centering
\includegraphics[width=\columnwidth]{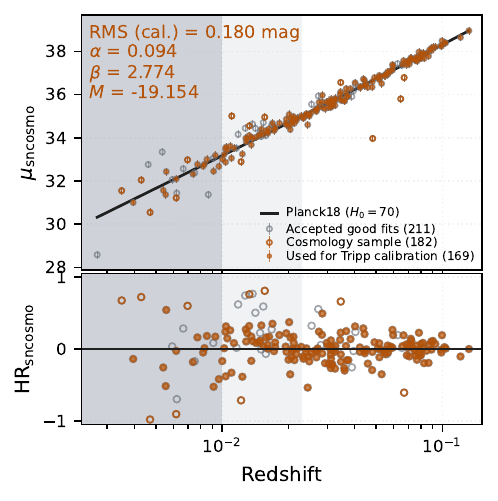}
\caption{Hubble-diagram for the final 211 SNe~Ia sample used in this work. The \texttt{sncosmo} SALT3-NIR Tripp-like standardization parameters are measured from a final 169-object subset, obtained after the standard light-curve quality cuts reduce the sample to 182 SNe~Ia and iterative $3\sigma$ clipping removes the remaining outliers. Open grey symbols mark \revdel{accepted good} \revadd{converged} fits not used for the cosmology calibration, open brown symbols mark objects in the 182-object cosmology sample but not in the final 169-object Tripp-calibration subset, and filled brown symbols mark the final Tripp-calibration sample. The reported RMS value corresponds to the 169-object calibration subset. The shaded regions indicate the low-redshift cuts, $z<0.01$ and $z<0.023$.}
\label{fig:hubble_diagram_comparison}
\end{figure}

\subsection{SN light-curve fitting}\label{sec:anasn}

The compiled light-curve sample is analyzed with \texttt{sncosmo} \citep{Barbary2016sncosmo} using the SALT3-NIR empirical model \citep{Pierel2022SALT3NIR}. We correct for Milky Way foreground reddening using the SN coordinates, the \citet{Schlafly2011} dust maps, and a \citet{Fitzpatrick1999} extinction law. Because SALT3-NIR extends the empirical model into the near-infrared (NIR), the retained optical+NIR information can be used more directly. 
\revdel{With this broader wavelength coverage, all 211 SNe~Ia yield acceptable SALT3-NIR fits.}
\revadd{With this broader wavelength coverage, all 211 light curves that pass the quality requirements of Sect.~\ref{sec:lcdata} yield converged SALT3-NIR fits, defined as the minimiser reaching a physical solution with finite parameter uncertainties and none of $x_0$, $x_1$, or $c$ pinned at the model bounds. We do not impose a reduced-$\chi^2$ cut: because the photometry is compiled from heterogeneous sources whose per-epoch uncertainties are not homogeneously calibrated, the reduced $\chi^2$ is dominated by error-scale mismatches rather than genuine fit quality and is not a reliable indicator. The samples are instead defined by the data-level requirements above and by the standard SALT parameter cuts below together with the iterative $3\sigma$ clipping.}
For each successful \texttt{sncosmo} fit we recover the standard $(x_0, x_1, c)$ parameters and convert the fitted amplitude to an effective rest-frame peak $B$-band magnitude through $m_B = - 2.5\log_{10}(x_0) + 10.6133$, where the constant corresponds to the $B$ peak magnitude in the Vega system for a $(x_0,x_1,c,z)=(1.0,0.0,0.0,0.0)$ in the SALT3-NIR model. The distance modulus is determined using the \cite{1998A&A...331..815T} relation, 
\begin{equation}
\mu_{\mathrm{SN}} = m_B + \alpha x_1 - \beta c - M. 
\end{equation}
To determine the nuisance parameters $(\alpha,\beta,M)$, we start from the full set of 211 \revdel{accepted} \revadd{converged} SALT3-NIR light-curve fits and apply the standard SALT light-curve-quality cuts, $-3 < x_1 < 3$ and $-0.3 < c < 0.3$. This leaves 182 SNe for the calibration sample. We then perform an iterative 3$\sigma$ clipping on the resulting Hubble residuals (HRs), removing outliers until convergence and yielding a final 169-object subset. On this clipped sample, $\alpha$, $\beta$, and $M$ are inferred with a Bayesian fit to the Tripp relation, simultaneously solving for an intrinsic-scatter term $\sigma_{\rm int}$, using the affine-invariant Markov chain Monte Carlo sampler implemented in \texttt{emcee}. We use 32 walkers evolved for 2500 steps each, discard the first 500 steps of every chain as burn-in, and adopt the posterior medians as the baseline calibration, with the 16th and 84th percentiles defining the corresponding uncertainties. The resulting parameters are $\alpha = 0.094^{+0.012}_{-0.012}$, $\beta = 2.774^{+0.164}_{-0.161}$, $M = -19.154^{+0.015}_{-0.014}$ mag, and $\sigma_{\rm int} = 0.182$ mag. These calibrated coefficients are then used to compute standardized distance moduli, $\mu_{\mathrm{SN}}$, and Hubble residuals $\mathrm{HR} = \mu_{\mathrm{SN}} - \mu_{\mathrm{cosmo}}(z_{\rm CMB})$\revadd{, so that a positive (negative) HR corresponds to a SN that is fainter (brighter) than the cosmological model predicts after standardization}. Figure~\ref{fig:hubble_diagram_comparison} shows the \texttt{sncosmo} SALT3-NIR Hubble diagram and HRs for the full sample after applying each cut. For the calibration sample (169 SNe Ia), the HR scatter is $\mathrm{RMS}(\mathrm{HR}_{\mathrm{sncosmo}})=0.180$~mag. Applying redshift cuts reduce the RMS to 0.166~mag for $z>0.01$ (156 SNe Ia) and 0.151~mag for $z>0.023$ (109 SNe Ia). 

%%%%%%%%%%%%%%%%%%%%%%%%%%%%%%%%%%%%%%%%%%%%%%%%%%%%%%%
%%%%%%%%%%%%%%%%%%%%%%%%%%%%%%%%%%%%%%%%%%%%%%%%%%%%%%%
%%%%%%%%%%%%%%%%%%%%%%%%%%%%%%%%%%%%%%%%%%%%%%%%%%%%%%%

\section{Results}

%%%%%%%%%%%%%%%%%%%%%%%%%%%%%%%%%%%%%%%%%%%%%%%%%%%%%%%

\subsection{Properties of the local UV environment}\label{sec:uv}

\begin{figure}[!t]
\centering
\includegraphics[width=\columnwidth]{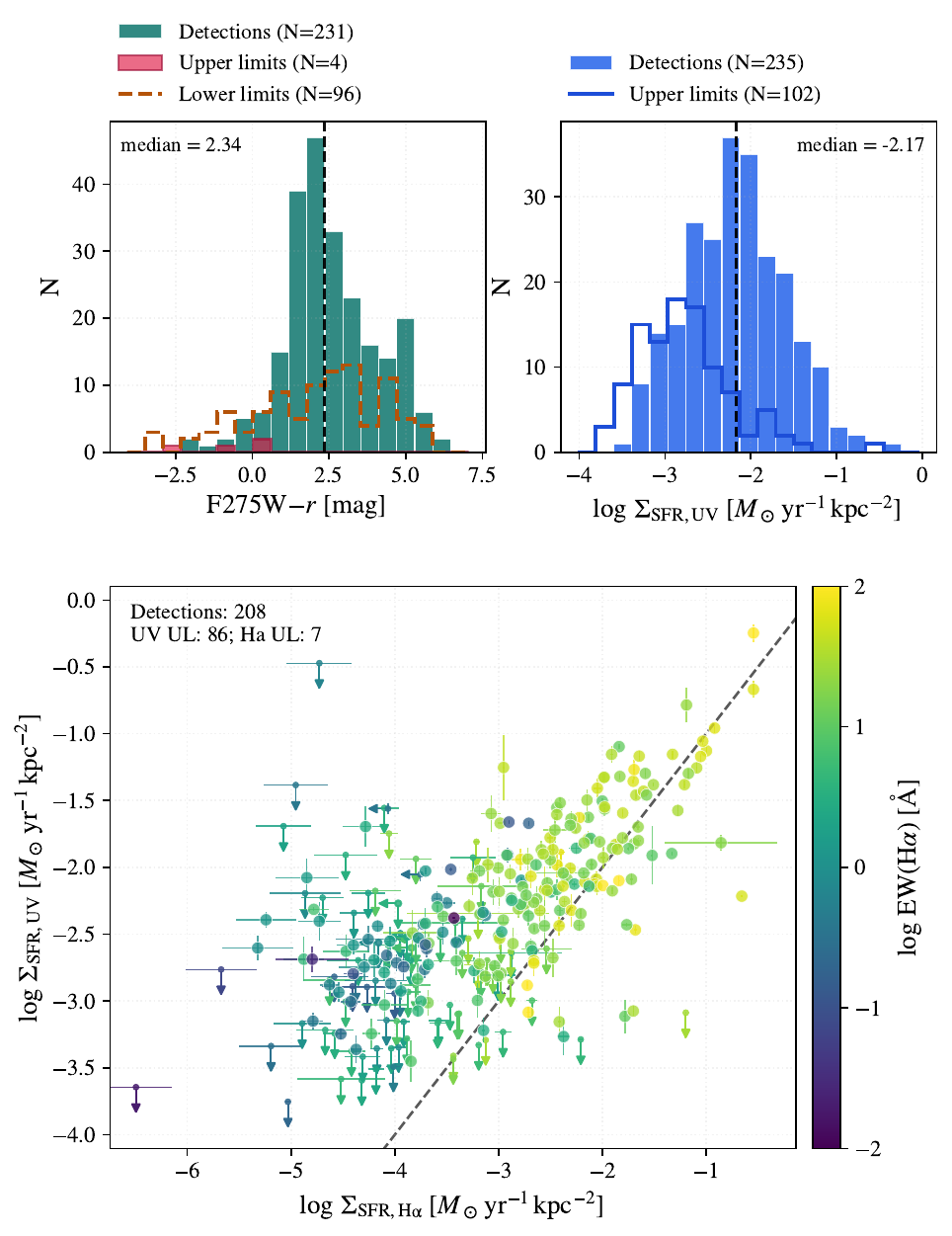}
\caption{Distributions and cross-comparison of local UV- and H$\alpha$-based star-formation indicators. Top-left: host-corrected ${\rm F275W}-r$ distribution, with detections (filled green), upper limits (filled magenta), and lower limits (brown dashed). Top-right: $\log \Sigma_{\rm SFR,UV}$ distribution, with detections (filled blue) and upper limits (blue outline). Bottom: $\log\,\Sigma_{\rm SFR,UV}$ versus $\log\,\Sigma_{\rm SFR,H\alpha}$ measured within the same 1~kpc aperture. Circles show UV detections while downward arrows mark UV upper limits, all colored by EW(H$\alpha$). The dashed line indicates the one-to-one relation. For the lower panel, 39 environments \revdel{classified as AGN in the BPT diagram} \revadd{with AGN-like line ratios in the [N\,{\sc ii}] BPT diagram (Sect.~\ref{sec:gas})} are excluded.}
\label{fig:uv_distributions_sfr_compare}
\end{figure}

Of the 337 host galaxies with valid HST F275W imaging, 235 (70\%) yield significant UV detections within the 1 kpc aperture centred on the SN position, while 102 (30\%) provide upper limits that define the locally UV-passive subsample. For the host-corrected F275W$-r$ color, 231 environments have direct measurements, 4 provide upper limits, and 96 provide lower limits, while 6 lack a usable color estimate. The UV SFR surface density among detections spans more than three orders of magnitude, with a median of $\log\,\Sigma_{\rm SFR,UV} = -2.17$ M$_\odot$ yr$^{-1}$ kpc$^{-2}$ and a normalized median absolute deviation ($\mathrm{nMAD}$\footnote{$1.4826 \times \mathrm{median}\left(\left| x_i - \mathrm{median}(x) \right|\right)$.}) of $0.57$ dex. The F275W$-r$ color distribution extends from $-1.95$ to $+6.46$ mag, with a median of $2.34$ mag and $\mathrm{nMAD}=1.30$ mag, reflecting the full range of local environments from actively star-forming clumps to quiescent stellar populations. Both distributions are shown in the upper panels of Fig.~\ref{fig:uv_distributions_sfr_compare}, including detections and non-detections.

A comparison of our two SFR tracers is shown in the bottom panel of the same figure, which compares $\Sigma_{\rm SFR,UV}$ from F275W photometry with $\Sigma_{\rm SFR,H\alpha}$ from the MUSE spectrum in the same 1 kpc aperture. After excluding 39 environments \revdel{classified as AGN in the BPT diagram} \revadd{with AGN-like line ratios in the [N\,{\sc ii}] BPT diagram (Sect.~\ref{sec:gas}; composite environments are retained)}, 296 galaxies remain in the comparison sample, including 208 detections in both tracers, 81 systems with only a UV upper limit, 2 with only an H$\alpha$ upper limit, and 5 with limits in both quantities. For the 208 galaxies detected in both tracers, the two are strongly correlated (Pearson $r = 0.690$, $p = 1.1\times10^{-30}$), confirming that both tracers probe similarly the underlying star formation activity. However, the UV-derived SFR exceeds the H$\alpha$-derived value by a median of $0.50$ dex. 
\revdel{This offset is physically expected and reflects the different timescales probed by the two indicators: H$\alpha$ emission traces ionising stars with ages $\lesssim 10$ Myr, whereas the F275W continuum integrates star formation over $\sim 100$--$200$ Myr. Environments with high UV-to-H$\alpha$ ratios have therefore experienced sustained star formation over the past few $10^8$ yr that is now declining, while systems lying closer to the one-to-one relation are more consistent with steady or rising recent star formation.} 
\revadd{A median UV excess of this amplitude is consistent with the different timescales probed by the two indicators: H$\alpha$ traces ionizing stars with ages $\lesssim 10$ Myr, whereas the F275W continuum integrates star formation over $\sim 100$--$200$ Myr. An offset of this size is not uniquely predicted by the timescale difference alone, and may also reflect differential dust attenuation between the nebular and stellar components, partial escape or absorption of ionising photons, stochastic IMF sampling at low SFR surface densities, and the differing assumptions of the two SFR calibrations. Elevated UV-to-H$\alpha$ ratios are thus qualitatively consistent with a recent decline in star formation, and near one-to-one ratios with steady or rising recent star formation, but a robust interpretation would require explicit star-formation-history modelling.} 
As we discuss in the following subsections, this timescale difference has direct physical implications for interpreting the connection between the local environment and SN\,Ia properties.

%%%%%%%%%%%%%%%%%%%%%%%%%%%%%%%%%%%%%%%%%%%%%%%%%%%%%%%

\subsection{The impact of UV data on stellar population fitting}

A central motivation of this work is the expectation that UV photometry helps break the age--metallicity--reddening degeneracy inherent to optical-only stellar population fitting. We can test this directly by comparing the \texttt{STARLIGHT} results from the optical-only fit with those from the joint UV+optical fit for the same 334 galaxies where both are available.

\begin{figure}[!t]
\centering
\includegraphics[width=\columnwidth]{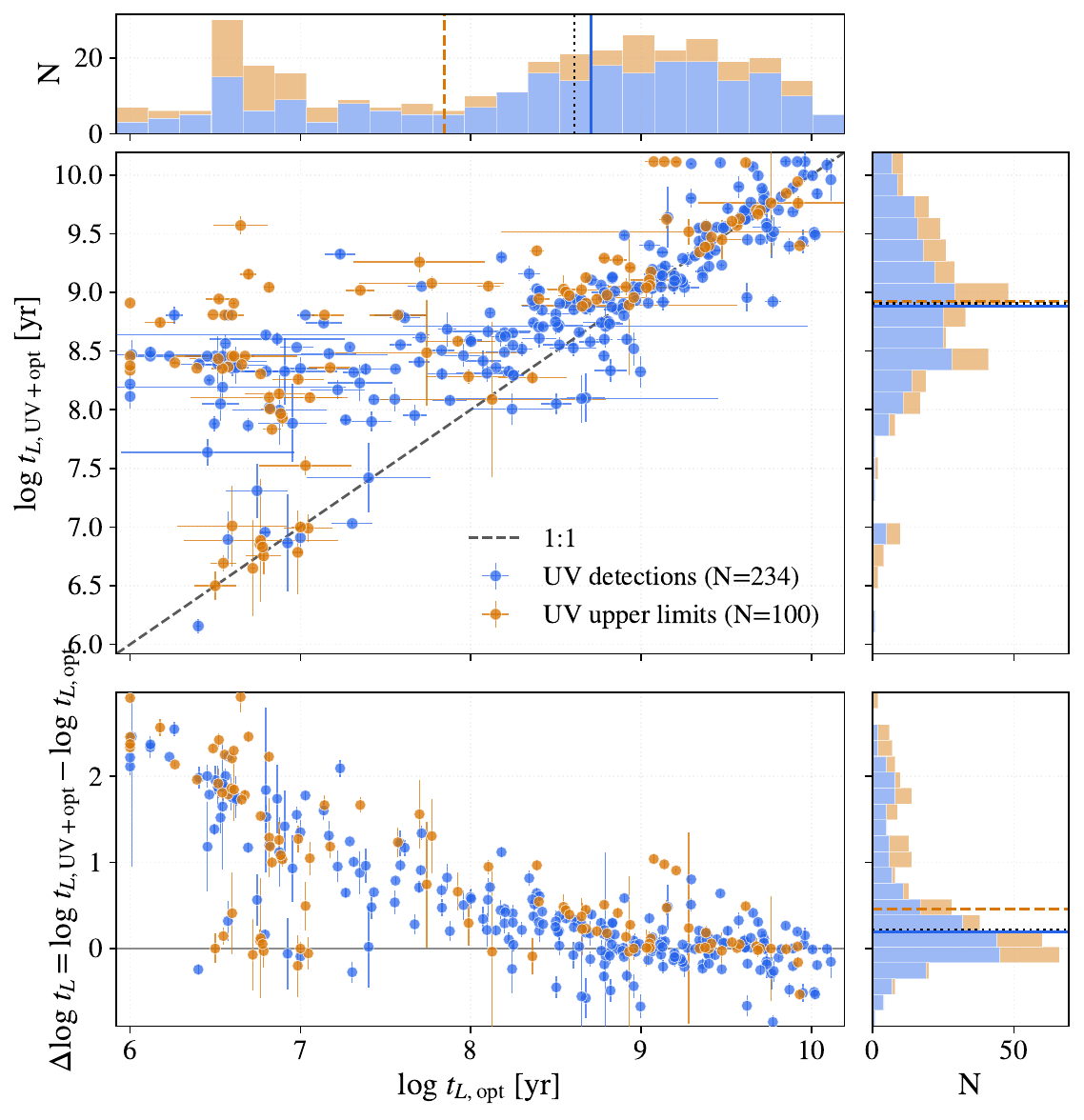}
\caption{Comparison of local luminosity-weighted stellar ages from \texttt{STARLIGHT} optical-only and joint UV+optical fits for the 334 galaxies with both measurements available. Top panel: $\log\,t_{L,\mathrm{UV+opt}}$ versus $\log\,t_{L,\mathrm{opt}}$, color-coded by UV detections (blue; $N=234$) and UV upper limits (brown; $N=100$) used in the \texttt{STARLIGHT} fit. The diagonal dashed line marks the one-to-one relation. The top and right panels show the stacked distributions of $\log\,t_{L,\mathrm{opt}}$ and $\log\,t_{L,\mathrm{UV+opt}}$. Bottom panel: $\Delta\log t_L \equiv \log\,t_{L,\mathrm{UV+opt}}-\log\,t_{L,\mathrm{opt}}$ as a function of $\log\,t_{L,\mathrm{opt}}$, with the right marginal panel showing the stacked distribution of $\Delta\log t_L$. In all distribution panels, vertical/horizontal lines mark the medians of UV detections (blue solid), UV upper limits (brown dashed), and the combined sample (black dotted). \revadd{The short vertical ticks at the bottom of the top panel mark the 15 \citet{BruzualCharlot2003} SSP age-grid points ($\log t/\mathrm{yr}=6.0$--$10.1$).}}
\label{fig:logt_uvopt_vs_opt}
\end{figure}

The most striking effect of including the UV constraint is on the luminosity-weighted stellar age, $\log\,t_L$. For UV-detected galaxies, the joint fit yields ages that are a median of $+0.19$ dex older than the optical-only estimate, consistent with the UV anchoring the young stellar component and reducing ambiguity in the age--extinction degeneracy. The effect is far more pronounced for galaxies with UV upper limits, for which the median shift reaches $+0.46$ dex toward older ages relative to the optical-only estimate. \revadd{This shift is robust to the adopted upper-limit level: re-fitting with the F275W limits redefined at 1$\sigma$, 2$\sigma$ and 5$\sigma$ changes the median joint-fit age of the non-detection population by $\lesssim0.2$~dex across the full range ($\log t_L=8.8$--$9.0$, with deeper limits giving slightly older ages), and the upward revision relative to the optical-only ages persists at every level (median $+0.29$ to $+0.91$ dex).} As shown in Fig.~\ref{fig:logt_uvopt_vs_opt}, the two fits are broadly consistent for systems older than a few $10^8$ yr, but at younger inferred ages the optical-only solution can move to values up to nearly two orders of magnitude younger. This systematic shift arises because, without a UV anchor, the optical spectrum alone can be fit with a young stellar component that remains formally acceptable in the optical but is ruled out by the UV non-detection. Once the UV constraint is included, that young component is suppressed and the inferred stellar population shifts to substantially older ages. Overall, 74.6\% of the sample has a joint-fit age older than the optical-only age, with differences exceeding 0.5 dex in 35.3\% of galaxies and exceeding 1.0 dex in 24.3\%. \revadd{Crucially, this upward revision is not confined to the UV non-detections. Among UV detections, for which the F275W constraint is a two-sided measurement, the joint-fit age is older than the optical-only value in 71\% of cases (median $+0.19$ dex), and the young optical-only detections ($\log t_{L,\rm opt}<7.5$) shift upward by a median of $+1.6$ dex, landing at essentially the same joint-fit age ($\log t_L\approx8.3$) as the corresponding non-detections. The effect therefore reflects the genuine UV constraint on the young stellar population rather than the one-sided treatment of the upper limits.} \revadd{These large shifts are not an artefact of regression toward the edges of the age grid ($\log t/\mathrm{yr}=6.0$--$10.1$; Sect.~\ref{sec:sp}): the luminosity-weighted age is a light-weighted mean over the SSP components and varies continuously between the grid limits, and the objects that shift the most land at intermediate joint-fit ages ($\log t_L\approx8.3$, range $6.2$--$9.6$), well inside the grid rather than piling up at the old boundary.}

This result has a practical consequence: the optical-only fit \revdel{artificially compresses} \revadd{tends to compress} the age distribution of locally passive galaxies toward younger values, diluting any age-based signal. The joint fit \revdel{recovers a more physically plausible age distribution} \revadd{shifts this distribution toward older ages and tightens it}, although for UV non-detections the constraint comes from the absence of UV emission rather than from a direct detection. By contrast, the mass-weighted age $\log\,t_M$ is much less affected (median shift $+0.03$ dex), as expected because it is dominated by the old stellar population that contributes little UV flux. Stellar masses are likewise largely unchanged between the two fits (median shift $+0.03$ dex, $\sigma = 0.33$ dex).

\begin{figure}[!t]
\centering
\includegraphics[width=\columnwidth]{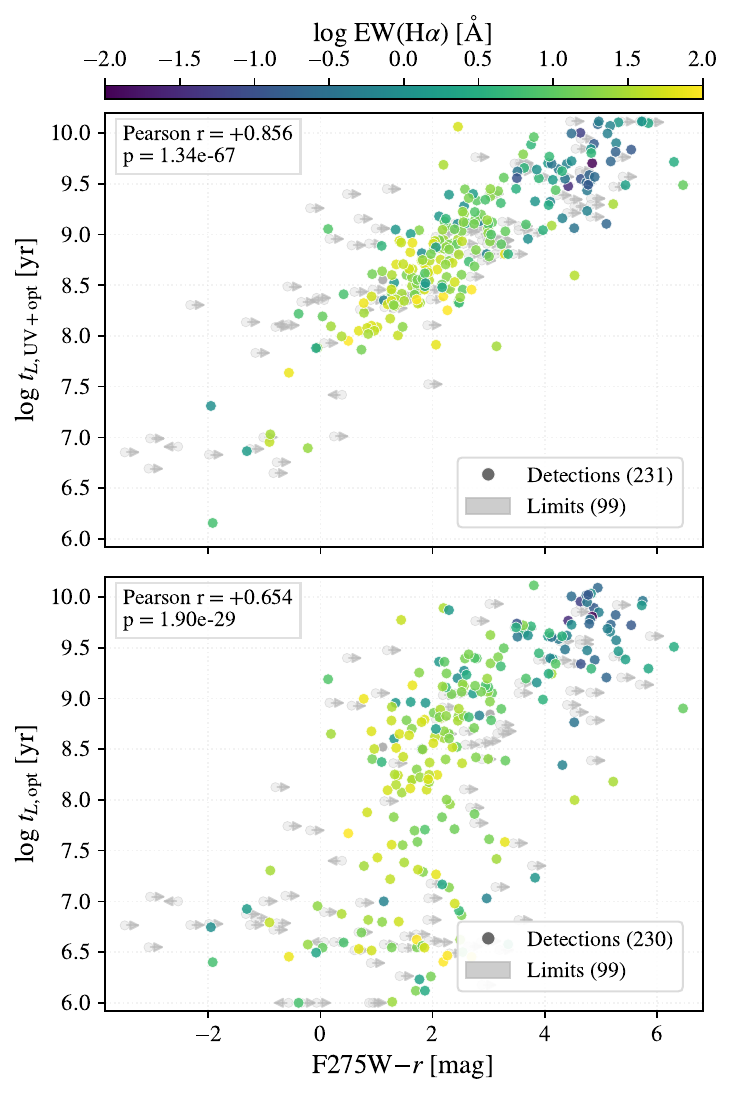}
\caption{Comparison of luminosity-weighted stellar age versus host-corrected F275W$-r$ color for the UV+optical fit (top) and the optical-only fit (bottom).
Galaxies with measured F275W$-r$ are shown as filled circles and are color-coded by $\log\,\mathrm{EW}(\mathrm{H}\alpha)$. F275W$-r$ upper and lower limits are shown with horizontal arrows in light gray. Pearson $r$ and $p$ values are computed using the color detections only.}
\label{fig:f275w_r_logt_compare_ewha}
\end{figure}

%%%%%%%%%%%%%%%%%%%%%%%%%%%%%%%%%%%%%%%%%%%%%%%%%%%%%%%

\subsection{UV color as a direct tracer of local stellar age}\label{sec:uvcol}

Figure \ref{fig:f275w_r_logt_compare_ewha} illustrates how the inferred local stellar age varies with host-corrected F275W$-r$ color, comparing the joint UV+optical fit with the optical-only fit. The separation between blue, UV-bright environments and redder, UV-passive ones is much clearer for the UV+optical fit ages, showing directly that the UV constraint does not simply regularize an underconstrained optical solution, but actively reshapes the recovered stellar populations.

For the joint fit, F275W$-r$ correlates very strongly with the luminosity-weighted age ($r = 0.856$, $p = 1.3\times10^{-67}$), as expected if the UV$-$optical color is tracing the fraction of young stars that dominates the recent star-formation history. Because both F275W$-r$ and $t_{L,\mathrm{UV+opt}}$ are constrained by the same UV information, we interpret this tight relation primarily as an internal consistency check of the joint fit rather than as an independent validation. By contrast, the optical-only age shows a substantially weaker, though still significant, correlation with the same color ($r = 0.654$, $p = 1.9\times10^{-29}$), indicating that the UV anchor reduces the age--metallicity--reddening degeneracy and \revdel{yields a more reliable local age estimator for} \revadd{tightens the inferred local age used in} the subsequent SN-environment analysis. \revadd{Because the optical-only age uses no F275W information, this weaker but highly significant correlation provides an F275W-independent check that the color traces stellar age and substantially reduces the shared-input covariance noted above. The host-corrected color used here still carries some joint-fit information through its $A_V^\star$ attenuation correction. Repeating the test with the Milky-Way-corrected only F275W$-r$ color, which uses no \texttt{STARLIGHT} attenuation at all, yields an essentially identical correlation with the optical-only age ($r=0.667$, $p=5.5\times10^{-31}$), and adopting instead an attenuation from the optical-only fit gives $r=0.774$, so the check is not driven by shared inputs.}

The behaviour of the color limits is also informative. Most non-detections correspond to lower limits on F275W$-r$, that is, to environments that are redder than the measured detections. These systems preferentially populate the old-age end of the joint-fit relation, consistent with the idea that the UV upper limits suppress spurious young components in the stellar-population fit.

The mass-weighted age from the joint fit also correlates with UV$-r$, but more weakly than the luminosity-weighted age ($r = 0.611$, $p = 5.2\times10^{-25}$), as expected given that it is dominated by the old stellar component. By contrast, the correlation with gas-phase metallicity is weak and not statistically significant ($r = 0.068$, $p = 0.304$), confirming that F275W$-r$ functions primarily as an age indicator rather than a metallicity tracer in our sample. This is physically consistent with theoretical stellar population models: at the ages and metallicities relevant here, the UV$-$optical color is governed mainly by the fraction of young, hot stars and is relatively insensitive to metallicity variations of the amplitude present in our sample \citep{2003MNRAS.342..259D,2013ARA&A..51..393C}. Taken together, these results \revdel{establish F275W$-r$ as a reliable photometric proxy for} \revadd{support the use of F275W$-r$ as an empirical photometric tracer of} the local stellar population age at 1~kpc resolution, \revdel{directly probing} \revadd{statistically related to} the timescales most relevant to the SN\,Ia delay-time distribution.

%%%%%%%%%%%%%%%%%%%%%%%%%%%%%%%%%%%%%%%%%%%%%%%%%%%%%%%

\subsection{\revadd{SN\,Ia local environment correlations}}\label{sec:lcenv}

We now connect the local environment directly to the SN\,Ia observables, focusing on the light-curve stretch parameter $x_1$, the color parameter $c$, \revadd{and the Hubble residuals $HR$} from the \texttt{sncosmo} SALT3-NIR fits. The analysis in this subsection is restricted to the 169 SNe Ia used for the Tripp calibration (Section~\ref{sec:anasn}), which pass the standard light-curve quality cuts and the iterative $3\sigma$ clipping, so that all correlations are evaluated on the same well-behaved, cosmology-ready subsample. \revdel{Table~\ref{tab:env_sncosmo_median_split_summary_calibration} reports the Pearson correlation coefficients, $p$-values, and median values of each bin and of the differences for all environmental parameters against $x_1$ and $c$, with each sample split at the median of the environmental variable. This two-bin diagnostic is less sensitive to outliers than a full linear fit and is directly comparable to the step-based formalism used in previous local-environment SN\,Ia studies.}

\revadd{We characterize each environment--SN relation with two complementary \emph{bivariate} diagnostics, which relate one environmental tracer to one SN observable at a time, and then assess the tracer set as a whole with a \emph{multivariate} analysis:}\\
\revadd{(i) The \emph{two-bin} split by the median of the environmental variable is the step-based diagnostic used in previous local-environment SN\,Ia studies, and it is less sensitive to outliers than a full linear fit.  We quote the offset between the low- and high-$x$ bin medians and its significance $\sigma$. Its robustness is tested in two ways. First, the dividing threshold is scanned over the 30th--70th percentile of the environmental variable. Second, for the three UV-based tracers with non-detections, a censored Monte Carlo resamples each limit $L$ beyond its value and recomputes the split: down to a floor $f$ placed $0.5$~dex below the smaller of the first percentile of the detections and the lowest limit (for the upper-limited sSFR and $\Sigma_{\rm SFR}$), or up to $0.5$~mag above the reddest detection (for the lower-limited F275W$-r$). The replacements are drawn under three prescriptions: uniform in the log tracer [$\mathcal{U}(f,L)$], uniform in the linear quantity [$\log_{10}\mathcal{U}(10^{f},10^{L})$], and a Gaussian $\mathcal{N}(\mu,\sigma)$ truncated to the interval.}\\
\revadd{(ii) In the \emph{linear} view we treat the two quantities as continuous over the $N_{\rm det}$ detections. We quote the Pearson $r$, Spearman $\rho$ and Kendall $\tau$ coefficients, each with its two-sided $p$-value. These latter two rank statistics are robust to non-linearity and to a few extreme measurements. Because each SN parameter is tested against twelve tracers simultaneously, we also apply a \cite{Benjamini95} false-discovery-rate (FDR) correction to each family of twelve Spearman $p$-values. In practice, the twelve $p$-values are ranked in ascending order, rescaled by $n/k$ ($n=12$, $k$ the rank), made monotonic from the largest downwards, and those with an adjusted $p<0.05$ are retained as significant. This leaves the smallest $p$-values essentially unchanged while strongly penalizing isolated marginal ones, so that the expected fraction of false positives among the relations we call significant is controlled. Using Kendall $\tau$ or Pearson $r$ instead leaves the set of surviving relations unchanged in all three cases ($x_1$, $c$ and HR). We also fit each relation with a maximum-likelihood linear regression that folds in the measurement errors in both variables and an intrinsic-scatter term, reporting the slope $b\pm\sigma_b$, its significance $|b/\sigma_b|$, and $\sigma_{\rm int}$. }\\
\revadd{(iii) Because the twelve tracers are far from independent, a \emph{multivariate} analysis assesses the tracer set as a whole. Their mutual dependences are quantified by the correlation matrix of the twelve tracers (Fig.~\ref{fig:tracer_corr}). The information carried individually by each tracer is then isolated with partial correlations: the SN observable and the tracer are each regressed on a set of control variables, and the correlation between the two sets of residuals measures the association that survives once the controls are accounted for. Finally, a standardized multiple regression of the SN observable on the tracers tests whether any single one retains an individually significant coefficient when all are fitted simultaneously. Because the surface-density and specific SFRs differ only by the stellar mass ($\log \Sigma$SFR$\,=\log$\,sSFR$\,+\log M+$\,const), the twelve-tracer design matrix is exactly singular. The two surface densities are therefore excluded and the regression is run on the remaining ten tracers (complete-case subset, $N=117$). The correlation matrix characterizes the tracer set itself, whereas the partial correlations and the multiple regression involve an SN observable and are computed separately for $x_1$, $c$ and HR.}

\revadd{Table~\ref{tab:env_sncosmo_median_split_summary_calibration} reports, for all environmental parameters against $x_1$, $c$ and HR (each split at the median of the environmental variable), the bin medians, their difference, and its significance, together with the Pearson, Spearman and Kendall correlation coefficients and the robustness diagnostics for the same sample.}
\revadd{The qualitative descriptors used throughout the correlation analysis are shorthand for these values (``very strong'': $|r|\gtrsim0.7$ or $\gtrsim7\sigma$; ``strong'': $\gtrsim5\sigma$; ``weak'' or ``marginal'': below $3\sigma$). We present these diagnostics for $x_1$ in Sect.~\ref{sec:lcenvx1}, for $c$ in Sect.~\ref{sec:lcenvc}, while the corresponding analysis of the Hubble residuals is given in Sect.~\ref{sec:hr}.}

\revadd{\subsubsection{Local environment and SN\,Ia light curve width}\label{sec:lcenvx1}}

\begin{figure*}[t]
\centering
\includegraphics[width=\textwidth]{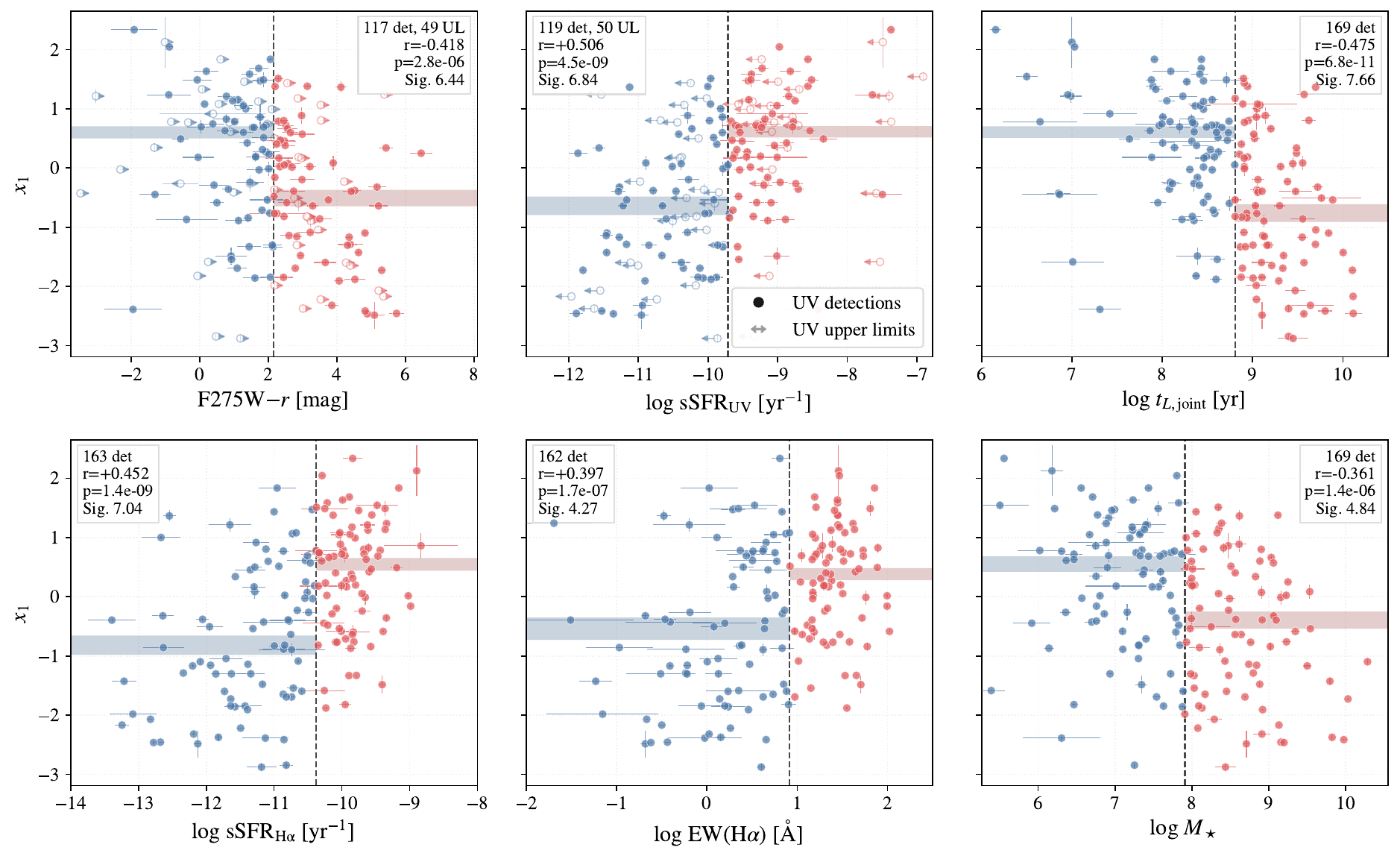}
\caption{Scatter plots between SN~Ia light-curve stretch ($x_1$) and local environment parameters from the UV+opt analysis: F275W$-r$, $\log {\rm sSFR_{UV}}$, $\log t_{L,{\rm joint}}$, $\log {\rm sSFR_{H\alpha}}$, $\log\,{\rm EW(H\alpha)}$, and $\log M_\star$. Points are split by the median of the x-variable (vertical dashed line), and colored in blue for the low-x bin and red for the high-x bin. In the UV-based panels, upper limits are shown with arrows. These non-detections are included both when defining the median x-split and in computing the low/high-bin $x_1$ medians. The blue/red horizontal bands indicate the bin medians in $x_1$ with their uncertainties. Each panel reports the number of detections/upper limits, the Pearson $r$ and $p$ values (from detections), and the significance of the bin difference, $\Delta x_1/\sigma_\Delta$. \revadd{The dotted horizontal line marks $x_1=0$.}}
\label{fig:x1_fourpanel_correlations}
\end{figure*}

Figure \ref{fig:x1_fourpanel_correlations} shows the six most relevant environment--$x_1$ diagnostics, covering both UV-based and IFS-based tracers: F275W$-r$, $\log {\rm sSFR_{UV}}$, $\log t_{L,{\rm joint}}$, $\log {\rm sSFR_{H\alpha}}$, $\log\,{\rm EW(H\alpha)}$, and stellar mass $\left( \log M_\star \right)$. All six panels show consistent trends in the expected direction: younger, more star-forming, and lower-mass local environments host SNe Ia with systematically broader light curves (higher $x_1$), while older, more quiescent, and more massive environments host narrower, faster-declining events.

\revdel{The hierarchy of predictors of $x_1$ is striking and reveals} \revadd{The individual median-split diagnostics reveal} the complementary role of the UV and IFS measurements. Among them, the joint-fit luminosity-weighted stellar age $\log t_L$ \revdel{is the strongest single predictor} \revadd{gives the largest nominal significance} ($\Delta x_1 = -1.37 \pm 0.18$; $7.66\sigma$), followed closely by the H$\alpha$-based local sSFR ($\Delta x_1 = 1.36 \pm 0.19$; $7.04\sigma$), the UV-based sSFR ($\Delta x_1 = 1.25 \pm 0.18$; $6.84\sigma$), and the F275W$-r$ color ($\Delta x_1 = -1.11 \pm 0.17$; $6.44\sigma$). \revdel{The H$\alpha$ sSFR and the UV sSFR therefore yield correlations with $x_1$ of very similar strength on the calibration subsample, in contrast to the more pronounced UV advantage found when using the full light-curve sample. This suggests that the additional scatter in the UV sSFR--$x_1$ relation at the faint end is partially absorbed by the cosmology-motivated light-curve cuts, and that when restricted to a cleaner sample both tracers carry similar information about the progenitor delay time.} The strong performance of F275W$-r$ despite relying only on local photometry reinforces its value as a \revdel{single-observable proxy for progenitor-age environment at 1 kpc scales} \revadd{single-observable tracer of the local stellar-population age at 1 kpc scales}.

The raw SFR surface densities remain weaker predictors than the corresponding specific SFR quantities, with the H$\alpha$-based surface density reaching only $2.29\sigma$ and the UV-based surface density showing a more modest but still non-negligible trend at $3.00\sigma$. \revdel{This confirms that sSFR is the more physically informative tracer of light-curve shape.} \revadd{This points to a stronger empirical association between the sSFR and light-curve shape than for the raw surface density.} Local stellar mass remains significant at $4.84\sigma$, and EW(H$\alpha$) yields $\Delta x_1 = 0.92 \pm 0.22$ ($4.27\sigma$). Gas-phase metallicity shows a weak but statistically significant anticorrelation with stretch ($r=-0.239$, $p=0.002$, $3.21\sigma$), broadly consistent with the expected mapping of the mass--metallicity relation onto the mass--stretch trend. These results confirm and extend the findings of \cite{2010MNRAS.406..782S} using global host properties, \cite{2013A&A...560A..66R} and \cite{2022A&A...657A..22B} using local IFS measurements, and are now established for the first time with a joint UV+IFS dataset at matched 1 kpc resolution. \revdel{The fact that the joint-fit stellar age anchored by the UV photometry emerges as the strongest predictor of $x_1$ directly validates} \revadd{That the joint-fit stellar age anchored by the UV photometry is among the leading predictors of $x_1$, on a par with the local sSFR, supports} the scientific motivation of the HST programme: the UV constraint provides age resolution in the regime of tens to hundreds of Myr that optical spectroscopy alone cannot deliver. The correlation of $x_1$ with the optical-only age is weaker ($4.2\sigma$), confirming that the improvement comes from the UV anchor rather than from the IFS data alone.

\revadd{The Spearman and Kendall rank coefficients track the Pearson values throughout, and the individually significant $x_1$ relations all survive the FDR correction across the twelve tracers. The maximum-likelihood regression recovers significant $x_1$ slopes for all the leading tracers ($4.0$--$6.3\sigma$; e.g.\ $\mathrm{d}x_1/\mathrm{d}\log t_L=-0.80\pm0.13$, $\sigma_{\rm int}\simeq1.0$), somewhat lower than the median-step significances because the continuous fit uses the full distribution rather than a two-bin contrast. The median-step significances are stable as the dividing threshold is moved over the 30th--70th percentile, and treating the UV upper limits as censored keeps the $x_1$--UV\,sSFR step highly significant ($4.8$--$5.8\sigma$). The same censored treatment of the F275W$-r$ color, whose UV non-detections are lower limits (redder than the limit), reduces its $x_1$ step from $6.4\sigma$ to $4.0\sigma$ but leaves it highly significant. The strong environment--$x_1$ relations are therefore robust to the correlation statistic, the multiplicity correction, the regression treatment, the split threshold and the censoring prescription.}

\revadd{A covariance and multivariate analysis further shows that these tracers do not act independently. On the calibration sample they are mutually correlated at $|r|$ up to $0.93$ (e.g.\ $\log t_L$ and UV\,sSFR, $r=-0.93$; H$\alpha$ sSFR and EW(H$\alpha$), $r=0.92$; Fig.~\ref{fig:tracer_corr}). A standardized multiple regression of $x_1$ on the ten tracers (complete-case subset, $N=117$) leaves none of the leading tracers individually significant. 
%Only the two stellar metallicities reach nominal significance, and they enter with opposite signs ($\beta=-0.44\pm0.19$ and $+0.41\pm0.20$) despite being mutually correlated at $r=0.91$, which is the signature of a pair of near-degenerate predictors rather than of an independent physical signal. The tracers are therefore largely redundant, and individual coefficients cannot be read as measures of relative importance.
 The nominal ranking among the leading tracers is likewise metric-dependent: the median split favors the joint age, whereas the Pearson correlation is marginally stronger for the UV\,sSFR ($|r|=0.51$ versus $0.48$). In partial-correlation terms $\log t_L$ and UV\,sSFR are largely interchangeable: controlling for the UV\,sSFR removes the age correlation entirely (partial $r=-0.04$, $p=0.69$), while controlling for the age leaves only a marginal UV\,sSFR signal (partial $r=+0.19$, $p=0.047$), and controlling jointly for the remaining leading tracers renders both insignificant. By contrast, $\log t_L$ remains a significant predictor when controlling separately for stellar mass, mass-weighted age, or F275W$-r$ (partial $r\simeq-0.26$ to $-0.33$), and the instantaneous H$\alpha$ sSFR retains a marginal independent signal beyond $\log t_L$ (partial $r=+0.27$, $p=0.004$). }
 
\revadd{The picture that emerges is therefore one in which $x_1$ is governed by a single dominant local ``young-stellar-fraction'' empirical axis, best captured by the joint UV+optical age or the UV sSFR, rather than with any uniquely identifiable parameter. The near-equal H$\alpha$ and UV sSFR significances on the calibration sample therefore reflect this shared dimension rather than a robust hierarchy, both tracers carrying similar information on the local stellar-population age and hence, in a statistical sense, on the progenitor delay times; the more pronounced UV advantage found in the full light-curve sample is consistent with the additional scatter at the faint end of the UV\,sSFR--$x_1$ relation being partially absorbed by the cosmology-motivated light-curve cuts.}

\revadd{\subsubsection{Local environment and SN\,Ia color}\label{sec:lcenvc}}

\begin{figure*}[t]
\centering
\includegraphics[width=\textwidth]{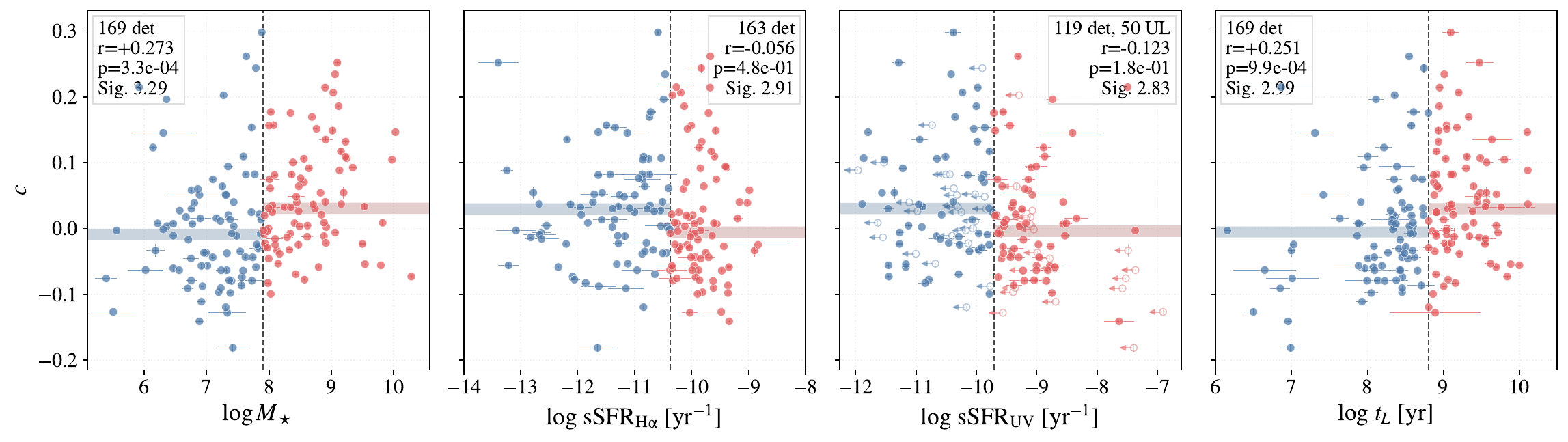}
\caption{Diagnostics for the \texttt{sncosmo} color parameter $c$ versus local environment properties, restricted to the 169 SNe Ia used in the Tripp calibration. The four panels show $c$ as a function of $\log M_\star$, $\log\,{\rm sSFR}_{\rm H\alpha}$, $\log\,{\rm sSFR}_{\rm UV}$, and $\log\,t_L$ (from the UV+optical fit). In each panel, points are split at the median of the x-variable; blue and red horizontal bands indicate the low- and high-bin medians of $c$ with their uncertainties, and the dashed vertical line marks the x-median split. For UV-based quantities, upper limits are shown with arrows and are included, using their limit values, both in defining the x-median split and in computing the low/high-bin medians. The annotation reports the number of detections/upper limits, Pearson correlation coefficient $r$, $p$-value (from detections), and the significance of the low/high-bin difference. \revadd{The dotted horizontal line marks $c=0$.}}
\label{fig:c_1x4_correlations_cal}
\end{figure*}

Turning to the SN color parameter $c$, Figure~\ref{fig:c_1x4_correlations_cal} shows four representative environment diagnostics spanning the strongest and most physically relevant trends. The correlations are weaker than for $x_1$, which is expected given that $c$ blends intrinsic SN color with line-of-sight dust. The strongest trends are with local stellar mass ($\Delta c = 0.040 \pm 0.012$; $3.29\sigma$), mass-weighted age ($\Delta c = +0.038 \pm 0.012$; $3.23\sigma$), light-weighted age ($\Delta c = +0.035 \pm 0.012$; $2.99\sigma$), H$\alpha$-based sSFR ($\Delta c = -0.036 \pm 0.012$; $2.91\sigma$), and UV-based sSFR ($\Delta c = -0.035 \pm 0.012$; $2.83\sigma$), with redder SNe Ia preferentially occurring in more massive, older, and less actively star-forming environments. These trends confirm, with matched 1 kpc measurements, the result of \cite{2010MNRAS.406..782S} and \cite{2013ApJ...770..108C} using integrated host properties, that SNe Ia in older, more massive, less star-forming environments are intrinsically or observationally redder. By contrast, EW(H$\alpha$) and the raw SFR surface densities show no strong correlation with $c$, and the F275W$-r$ color reaches only $2.35\sigma$\revdel{, consistent with the SN color parameter being driven by a physical mechanism (intrinsic color variation or dust) that is only partially traced by the local star formation activity}.

\revadd{The same checks applied to the SN color $c$ confirm that its environmental trends are genuine but secondary. The Spearman and Kendall coefficients again track the Pearson values (Table~\ref{tab:env_sncosmo_median_split_summary_calibration}), and under the FDR correction four of the twelve $c$ relations survive: local stellar mass, the light- and mass-weighted ages, and gas-phase metallicity. However, the sSFR- and SFR-based $c$ trends do not. The maximum-likelihood regression returns significant $c$ slopes for stellar mass ($3.5\sigma$), the two ages ($\sim3.3\sigma$) and metallicity ($2.7\sigma$), all weaker than the corresponding $x_1$ slopes, and the median-step significances are stable across the 30th--70th-percentile threshold scan. Treating the UV upper limits as censored leaves the already weak $c$--UV\,sSFR step below $1.3\sigma$.} \revadd{The multivariate analysis moreover identifies a single independent driver for $c$. In a standardized multiple regression on the ten tracers (complete-case subset, $N=117$), gas-phase metallicity is the only tracer with an individually significant coefficient ($\beta=+0.42\pm0.11$, $p=10^{-4}$), and it likewise retains the only significant partial correlation once the other leading tracers are controlled for (partial $r=+0.36$, $p=10^{-4}$), whereas stellar mass, both ages and the sSFRs lose their individual significance. }

\revadd{The environmental dependence of $c$ thus resides in the integrated properties of the local stellar population rather than in its instantaneous star-formation activity: EW(H$\alpha$) and the raw SFR surface densities show no significant trend, the sSFR-based trends do not survive the multiplicity correction, and the surviving mass and age trends are themselves largely absorbed by the local metallicity. This is consistent with the SN color being set by a mixture of intrinsic color variation and dust that follows the more metal-rich, more massive stellar component, with the mass--metallicity relation linking the two.}

\revadd{\subsubsection{Local environment and Hubble residuals}\label{sec:hr}}

\begin{figure*}[t]
\centering
\includegraphics[width=\textwidth]{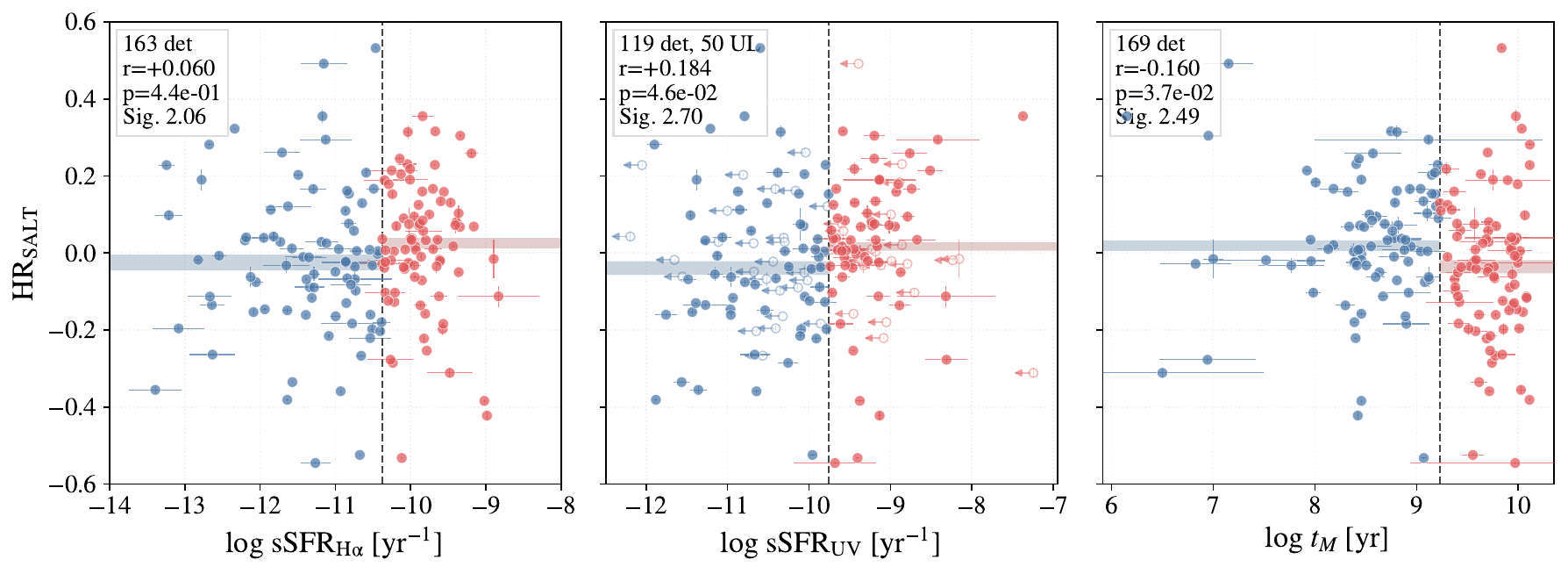}
\caption{Diagnostics for SALT Hubble residuals (${\rm HR}_{\rm SALT}$) versus local environment properties, restricted to the 169-object Tripp-calibration sample. The three panels show ${\rm HR}_{\rm SALT}$ as a function of $\log\,{\rm sSFR}_{\rm H\alpha}$, $\log\,{\rm sSFR}_{\rm UV}$, and $\log\,t_M$. In the UV-based panel, upper limits are shown with arrows and are included, using their limit values, both in defining the x-median split and in computing the low/high-bin medians. As in the other diagnostic figures, points are split at the median x-value, blue/red horizontal bands show low/high-bin central values with uncertainties, and each panel reports the number of detections/upper limits, Pearson $r$, $p$ (from detections), and the significance of the bin-to-bin difference. \revadd{The dotted horizontal line marks ${\rm HR}=0$.}}
\label{fig:hr_1x3_correlations_cal}
\end{figure*}

The Hubble residuals (HR~$\equiv \mu_{SN} - \mu_{\rm cosmo}$) reflect what remains of the SN luminosity scatter after stretch and color standardization, and represent the quantity of direct relevance to SN\,Ia cosmology.

Out of the twelve environmental quantities tested, none reach $3\sigma$ significance in the bin-to-bin difference after Tripp standardization. The strongest trend is with the mass-weighted stellar age, $\log t_M$, which shows $\Delta\mathrm{HR} = -0.055 \pm 0.022$ mag ($2.49\sigma$), such that older local environments host SNe Ia that are, on average, brighter after standardization. The local UV and H$\alpha$ sSFRs point in the same physical direction, with $\Delta\mathrm{HR} = +0.045 \pm 0.022$ mag ($2.09\sigma$) for the UV-based sSFR and $\Delta\mathrm{HR} = +0.050 \pm 0.024$ mag ($2.06\sigma$) for the H$\alpha$-based sSFR. Thus, even after stretch and color correction, SNe~Ia in locally younger, more actively star-forming regions tend to remain slightly fainter than those in older, more passive environments, although the evidence is only marginal on this calibration sample. Figure~\ref{fig:hr_1x3_correlations_cal} shows the HR dependence on the three environmental quantities for which the residual trends are largest.

All other environmental quantities, including the UV$-r$ color, the light-weighted age, EW(H$\alpha$), metallicity, and stellar mass, show no significant HR signal on this sample, as summarized in Table~\ref{tab:env_sncosmo_median_split_summary_calibration}. The sign of the three leading trends is broadly consistent with previous results on local environment steps \citep{2013A&A...560A..66R,2020A&A...644A.176R,2021A&A...649A..74N,2018A&A...615A..68R,2022A&A...657A..22B}. While this may seem counterintuitive given the established global mass step, a weaker local-$M_\star$ \revdel{signal} \revadd{trend} is not unexpected. Global stellar mass can encode host-wide information (e.g., metallicity, long-term star-formation history, and dust) that is only partially sampled by a fixed 1~kpc aperture. 

The absence of a significant UV$-r$--HR correlation, despite UV$-r$ being a strong predictor of both $x_1$ and $\log t_L$, is physically informative rather than contradictory. Most of the environmental imprint on the SN luminosity is already captured by the light-curve stretch: younger, more active environments produce broader, slower-declining SNe, which are then brought onto the standardised relation by the $\alpha x_1$ correction. The marginal UV sSFR \revdel{signal} \revadd{trend} that survives standardization points to a secondary effect not fully absorbed by stretch, and is consistent with the UV sSFR tracing \revdel{progenitor age} \revadd{the local stellar-population age (a statistical proxy for the progenitor age)} over longer timescales ($\sim 200$ Myr) than the instantaneous H$\alpha$-based tracers, and therefore probing a timescale closer to the full SN\,Ia DTD\revadd{. As shown below, however, this residual cannot be attributed to the UV\,sSFR specifically}.

\revadd{The same robustness checks applied to the $x_1$ and $c$ relations (Table~\ref{tab:env_sncosmo_median_split_summary_calibration}) confirm that these residual HR trends are only marginal. None of the twelve environmental tracers survives a FDR correction of the HR correlations (minimum FDR-adjusted $p=0.14$), and a maximum-likelihood regression that folds in the measurement errors and an intrinsic-scatter term yields HR slopes significant at only $\sim1.6\sigma$ (e.g.\ mass-weighted age and UV\,sSFR). Treating the UV upper limits as censored, further reduces the HR--UV\,sSFR step to $1.2$--$1.8\sigma$.}

\revadd{The multivariate analysis applied to the Hubble residuals identifies no independent driver either. In the standardized multiple regression on the ten tracers (complete-case subset, $N=117$) only the mass-weighted age and the mass-weighted metallicity reach nominal significance ($p=0.044$ and $0.029$), the latter again reflecting the near-degeneracy of the two stellar metallicities, and neither retains a significant partial correlation once the other leading tracers are controlled for (the largest being $\log t_M$, partial $r=-0.14$, $p=0.13$). The marginal HR trends of Table~\ref{tab:env_sncosmo_median_split_summary_calibration} therefore cannot be attributed to any individual environmental parameter: the tracers that carry them are the same covariant set that drives $x_1$, and the residual is most naturally read as a property of that shared axis, plausibly reflecting the longer timescales probed by the UV-based tracers, rather than as the signature of a specific quantity. Combined with the absence of any HR relation surviving the multiplicity correction and with regression slopes reaching only $\sim1.6\sigma$, this leads us to regard the post-standardization HR trends as exploratory rather than an established second environmental component, in contrast to the highly significant and robust environment--$x_1$ relations.}

We note that none of our environmental tracers individually reach the significance reported in previous studies that detected the local environment step. \cite{2015ApJ...802...20R} measured $\Delta{\rm HR} = 0.094 \pm 0.025$ mag at $\sim 3.7\sigma$ using GALEX FUV imaging within a 1 kpc aperture for 77 SNe Ia from the Constitution sample, and \cite{2020A&A...644A.176R} measured $\Delta{\rm HR} = 0.163 \pm 0.029$ mag at $5.7\sigma$ using spectroscopic H$\alpha$-based local sSFR within a 1 kpc aperture for 141 SNe Ia from the Nearby Supernova Factory, splitting the sample at $\log\mathrm{lsSFR} = -10.82$. The former is the most direct methodological comparison for our UV-based HR step, since both works classify the local environment from UV imaging at the same physical scale. Several factors plausibly contribute to the lower significance we obtain. First, our calibration sample is dominated by locally star-forming environments. The median local UV sSFR of our sample, used here as the bin-splitting threshold, is $\log\mathrm{lsSFR_{UV}} = -9.71$, more than one order of magnitude above the $\log\mathrm{lsSFR} = -10.82$ value adopted by \cite{2020A&A...644A.176R} to separate young from old environments. Adopting the Rigault cut on our sample shifts the split to a strongly imbalanced 28/137 (passive/active) configuration in UV and 57/102 in H$\alpha$, with the locally passive bin sparsely populated and individual outliers carrying significant weight. The corresponding step amplitudes would drop to $1.29\sigma$ (UV) and $0.99\sigma$ (H$\alpha$), compared to $2.09\sigma$ and $2.06\sigma$ at our median split. Second, the median redshift of our sample ($z = 0.028$) is lower than that of \cite{2020A&A...644A.176R} ($z = 0.032$), so peculiar-velocity uncertainties may contribute a non-negligible scatter to the residuals, which dilutes any small environmental step. Third, the heterogeneous origin of our light curves, assembled from CSP-I, CSP-II, ZTF, ATLAS, and the literature, introduces additional photometric scatter relative to the homogeneous SNfactory spectrophotometry of \cite{2020A&A...644A.176R}. Despite these caveats, the consistent direction and amplitude of the HR offsets across the three leading tracers (UV sSFR, mass-weighted age, and H$\alpha$ sSFR) still \revdel{support a residual environmental effect on} \revadd{are suggestive of a possible residual environmental trend in} standardised SN\,Ia luminosities at the $\sim 0.05$ mag level.

%%%%%%%%%%%%%%%%%%%%%%%%%%%%%%%%%%%%%%%%%%%%%%%%%%%%%%
%%%%%%%%%%%%%%%%%%%%%%%%%%%%%%%%%%%%%%%%%%%%%%%%%%%%%%
%%%%%%%%%%%%%%%%%%%%%%%%%%%%%%%%%%%%%%%%%%%%%%%%%%%%%%

\section{Discussion}

The results of Sections \ref{sec:uv}--\ref{sec:lcenv} support a coherent picture in which resolved UV information adds leverage that is not available from optical data alone, and therefore \revdel{validates}\revadd{supports} the strategy of combining HST UV imaging with matched-aperture IFS.

The strongest evidence comes from the close relation between F275W$-r$ and the joint-fit luminosity-weighted stellar age ($r = 0.856$), together with the substantial age revision introduced when the UV constraint is included in the \texttt{STARLIGHT} fits, particularly for UV-undetected environments. MUSE optical spectroscopy alone \revdel{systematically biases locally passive regions toward younger inferred ages} \revadd{tends to assign younger inferred ages to locally passive regions}, whereas the F275W detections and upper limits \revdel{suppress this degeneracy and recover a more plausible recent star-formation history} \revadd{restrict the allowed young stellar component and shift the inferred recent star-formation history accordingly}. In turn, the joint-fit luminosity-weighted age \revdel{is the single strongest predictor of $x_1$ in our sample} \revadd{gives the largest nominal median-split significance for $x_1$ in our sample}  ($r = -0.475$, $7.66\sigma$), closely followed by the H$\alpha$ sSFR and the UV sSFR \revadd{(the latter with a marginally stronger Pearson correlation)}.\revadd{ All the individually significant $x_1$ relations survive a false-discovery-rate correction applied across the twelve tracers.} The consistent \revdel{ranking} \revadd{behaviour} of these tracers indicates that \revadd{a single local ``young-stellar-fraction'' dimension, equivalently} \revdel{stellar age, or equivalently the fraction of young stars in the local progenitor environment,} \revadd{the local stellar age,} \revdel{is the main physical link between the environment and} \revadd{is the environmental parameter most strongly associated with} the SN~Ia light-curve width parameter\revadd{. Because the tracers are strongly covariant (Sect.~\ref{sec:lcenvx1}), we do not single out any one of them as uniquely fundamental}.

This interpretation is reinforced by the agreement between the UV and H$\alpha$ diagnostics. Although they probe different timescales, they recover local star formation in a highly consistent way and show comparable predictive power for $x_1$, while the joint UV+IFS fit performs slightly better than either tracer alone. The relation between UV- and H$\alpha$-based SFR surface densities, with UV exceeding H$\alpha$ by 
\revdel{about 0.36 dex on average, is naturally explained if many SN environments experienced stronger star formation over the last $\sim 200$ Myr than at the present epoch.} 
\revadd{a median of about 0.5 dex, is consistent with many SN environments having experienced stronger star formation over the last $\sim 200$ Myr than at the present epoch, although other effects can also contribute to the offset (Sect.~\ref{sec:uv}).} 
The combined dataset \revadd{may} therefore \revadd{help distinguish} not only whether a site is young or old, but also whether its recent star formation is ongoing or already declining.

The correlations involving SN color point in the same general direction, though with lower significance: redder SNe Ia are preferentially found in older, more massive, and less actively star-forming environments. \revadd{Of these, the trends with local stellar mass, stellar age and gas-phase metallicity are the ones that remain significant under a multiple-comparison (FDR) correction (Sect.~\ref{sec:lcenvc}), whereas the sSFR-based color trends are only marginal.} \revadd{A multivariate analysis further shows that, once the tracers are considered jointly, gas-phase metallicity is the only one that retains an independent association with $c$, the mass and age trends being largely absorbed by it (Sect.~\ref{sec:lcenvc}).} Because these trends are stronger for age- and \revdel{sSFR-based} \revadd{mass-based} quantities than for instantaneous tracers such as EW(H$\alpha$) or raw SFR surface density, they likely reflect integrated properties of the local stellar population rather than only the current level of star formation. This pattern is consistent with a mixture of intrinsic color differences and local dust effects.

After Tripp standardization, most of the environmental dependence is absorbed by the stretch correction, leaving only weak residual HR trends. The clearest remaining \revdel{signals are} \revadd{indications are} associated with UV sSFR and mass-weighted age, at the level of $\sim 0.05$ mag, but these are much less significant than the pre-standardization correlations with $x_1$\revadd{: no HR relation survives the same false-discovery-rate correction (minimum adjusted $p=0.14$), and the multivariate analysis identifies no individual tracer as an independent driver (Sect.~\ref{sec:hr})}. Overall, the \revdel{dominant environmental pathway appears to be the age--stretch relation}\revadd{dominant empirical association appears to be between stretch and the covariant young-stellar-population axis}: younger environments preferentially host broader SNe~Ia, while any additional dependence of standardized luminosity on the local environment is comparatively subtle. Even so, the fact that the residual trends are most evident in quantities informed by the UV data suggests that near-UV observations remain valuable for identifying and ultimately modeling environmental systematics in future precision SN~Ia samples.

These conclusions should, however, be interpreted with two methodological caveats in mind. First, the quantitative gain from adding F275W is partly data-dependent. At the low redshifts of our sample, MUSE does not cover the rest-frame $\sim 3600$--4160 \AA\ region, which contains the Balmer break and higher-order Balmer absorption lines and therefore provides important leverage against the age--dust--metallicity degeneracy. In that sense, the UV constraint in this work plays a dual role: it adds genuinely independent short-wavelength information, and it partly compensates for blue optical information that is absent from the MUSE wavelength range. Our comparison between optical-only and joint UV+optical fits should therefore be interpreted primarily in the context of nearby MUSE-based analyses, rather than as a universal calibration for all optical stellar-population studies, although previous work has likewise shown that UV and broader multiwavelength constraints can alter the recovered stellar-population properties \citep{2016MNRAS.458..184L,2019MNRAS.483..370B}. Second, our \texttt{STARLIGHT} fits adopt a single effective attenuation law for the full stellar-population mixture. Real galaxies are expected to show differential attenuation between young and old stellar populations, and because the UV is especially sensitive to dust, the inferred young stellar fractions, luminosity-weighted ages, and stellar attenuation values should be interpreted as effective parameters within this simplified framework rather than as a full description of the age-dependent dust geometry \citep{Calzetti2000,2020ARA&A..58..529S}.

A broader interpretative caveat concerns the physical meaning of our fixed 1~kpc aperture. Throughout this work, we treat the measured quantities as explosion-site environment proxies, recognizing that they may not perfectly coincide with the progenitor birth site because of stellar migration and finite delay times. The relevance of any fixed local aperture depends on the progenitor delay time and on internal galaxy dynamics: prompt channels should retain a stronger memory of their natal environment, whereas longer-delay channels may be increasingly decorrelated from it through orbital mixing and radial migration over roughly dynamical timescales \citep{2002MNRAS.336..785S}. For the star-forming disks that dominate our sample, a characteristic mixing timescale of order $\sim 100$ Myr makes 1~kpc a practical compromise: small enough to remain sensitive to local structure, but large enough to be robust against seeing, registration, and signal-to-noise limitations. In that sense, our measurements are best interpreted as physically motivated local proxies at a common scale, while a full multi-scale treatment of the delay-time distribution is deferred to future work. 
\revadd{We further stress that the local luminosity-weighted age is a light-weighted average over the stellar populations within the projected 1~kpc aperture, and not the age of the individual binary that produced the SN. The aperture can include unrelated populations along the line of sight, bulge and disc components, spiral-arm populations, or stars formed well before or after the progenitor, an effect that is especially severe in inclined and central regions. In particular, SNe~Ia in spirals also occur in the thicker, older stellar disc and at appreciable heights above the plane \citep{Hakobyan2017}, so the UV emission within the projected aperture may originate predominantly from young thin-disc stars while the progenitor belongs to an older thick-disc population. The UV-emitting population and the progenitor can therefore be physically unrelated even when they coincide on the sky. Luminosity-weighted, mass-weighted, and true progenitor ages also differ: at a fixed local population age the progenitor age follows a broad conditional distribution set by the local star-formation history and the SN~Ia delay-time distribution, so the local age constrains the progenitor age only in a statistical, ensemble sense rather than fixing the age of any individual progenitor \citep{2026arXiv260416597M,2026MNRAS.549ag797W}. We therefore use the local stellar-population age only as a statistical proxy for the progenitor-age distribution across the sample, rather than as the age of any individual system}.

A final caveat is that our environmental measurements are based on matched-aperture UV and optical data only, without the longer-wavelength near-, mid-, and far-infrared information that can help constrain dust attenuation, obscured star formation, and the cold-dust content of the host galaxies. In particular, infrared data can break degeneracies that remain even when UV constraints are available, by directly tracing the dust-reprocessed component of the stellar emission and by providing a more complete census of star formation in dusty environments. Recent SN~Ia host-galaxy studies have shown the value of combining optical data with broader-band SED information extending to the mid- and far-infrared, including \textit{Herschel}-based measurements, to derive more robust host stellar masses, star-formation rates, and evolutionary classifications \citep{2025MNRAS.543.2180R,2026MNRAS.549ag832R}. Extending the present resolved local-environment framework to include matched-aperture infrared constraints is therefore a natural next step, but is left for future work.

%%%%%%%%%%%%%%%%%%%%%%%%%%%%%%%%%%%%%%%%%%%%%%%%
%%%%%%%%%%%%%%%%%%%%%%%%%%%%%%%%%%%%%%%%%%%%%%%%
%%%%%%%%%%%%%%%%%%%%%%%%%%%%%%%%%%%%%%%%%%%%%%%%

\section{Summary and Conclusions}

We have assembled the largest matched-aperture ultraviolet and optical integral-field spectroscopy (UV+IFS) local-environment dataset for SN~Ia hosts to date, combining HST/WFC3 F275W imaging, ground-based $r$-band photometry, VLT/MUSE integral-field spectroscopy, and SN light-curve photometry. For each of the 340 SN~Ia host galaxies, all measurements are obtained within a common 1~kpc-radius aperture centered at the SN position, enabling a homogeneous analysis that links the local stellar population, ionised gas, and dust properties to the SN~Ia light-curve parameters and standardized luminosities. Light-curve fits with \texttt{sncosmo}/SALT3-NIR are available for 211 of the 340 SNe~Ia, and 169 of these pass the cosmology-motivated Tripp-calibration quality cuts used in the correlation analysis. Our main results are:

(i) \textit{UV detections and SFR tracers.} UV emission is detected at the SN site in 235 of the 337 environments with valid HST imaging ($\sim 70$\%), while the remaining 102 provide upper limits. The UV- and H$\alpha$-based SFR surface densities are strongly correlated, although the UV estimates are systematically higher by about 0.5 dex in the AGN-cleaned comparison sample, consistent with the different timescales traced by the two indicators ($\sim 200$ Myr and $\sim 10$ Myr, respectively). This offset likely reflects variations in the recent star-formation history at the SN site.

(ii) \textit{UV constraints \revdel{improve} \revadd{alter and tighten} local stellar-population ages.} Adding the F275W photometry to the joint \texttt{STARLIGHT} spectrophotometric fit shifts the inferred luminosity-weighted age $\log t_L$ upward by a median of $+0.20$ dex for environments with UV detections and $+0.45$ dex for those with UV non-detections, with shifts larger than 1 dex in 24\% of the sample. Optical-only fitting \revdel{systematically assigns artificially young ages to} \revadd{can assign younger ages to} locally passive environments by allowing \revdel{unconstrained }young stellar components that the UV upper limits \revdel{rule out} \revadd{restrict}. This effect is expected to be especially pronounced for nearby MUSE data, whose wavelength coverage lacks part of the blue optical age-sensitive region available in bluer optical surveys.

(iii) \textit{UV color as a local age indicator.} F275W$-r$ is very strongly correlated with the joint-fit luminosity-weighted age and shows no significant correlation with gas-phase metallicity, \revdel{demonstrating that it is a reliable empirical tracer of local stellar age on 1~kpc scales and an efficient single-observable proxy for progenitor age} \revadd{supporting its use as a statistical tracer of the local luminosity-weighted stellar-population age on 1~kpc scales within the adopted modelling framework, which serves as a statistical proxy for the progenitor age}.

(iv) \textit{Environment--stretch correlations.} On the 169-object Tripp-calibration sample, the joint-fit luminosity-weighted age \revdel{is the strongest predictor of SN~Ia stretch}\revadd{gives the largest nominal median-split significance for SN~Ia stretch}  ($r = -0.475$, $7.66\sigma$, $\Delta x_1 = -1.37 \pm 0.18$ across the median split), followed closely by H$\alpha$ sSFR and UV sSFR \revadd{(metric-dependent, since the UV sSFR has a marginally stronger Pearson correlation)}. The similar significance of the H$\alpha$- and UV-based sSFR measurements suggests that both trace the same underlying quantity: the fraction of young stars in the local progenitor environment. Across all tracers, younger and more actively star-forming environments host broader SNe~Ia, reinforcing the interpretation that \revdel{local stellar age is the main driver}\revadd{a single local ``young-stellar-fraction'' dimension---best traced by the joint UV+optical age and the local sSFRs, which are strongly covariant---is the dominant empirical axis associated with SN~Ia stretch}.

(v) \textit{SN color correlations.} The SALT3 color $c$ correlates \revdel{significantly}\revadd{most strongly} with local stellar mass, mass-weighted age, \revadd{and} light-weighted age\revdel{, H$\alpha$ sSFR, and UV sSFR}, such that redder SNe~Ia are preferentially found in older, more massive, and less actively star-forming environments\revadd{; these---together with gas-phase metallicity---are also the $c$ correlations that survive a false-discovery-rate correction, whereas the sSFR-based trends are only marginal}. By contrast, EW(H$\alpha$) and the raw SFR surface densities do not correlate significantly with $c$, suggesting that SN color is more sensitive to the integrated properties of the local stellar population than to the instantaneous star-formation activity. \revadd{When the tracers are analysed jointly, gas-phase metallicity is the only one that retains an independent association with $c$, indicating that the color--mass and color--age trends largely reflect the underlying mass--metallicity relation.}

(vi) \textit{Hubble residuals.} In the final 169-object calibration sample, no environmental parameter yields an HR signal above $3\sigma$. The strongest trend is associated with the mass-weighted age, followed by the UV- and H$\alpha$-based sSFRs, with amplitudes of order $\sim 0.05$ mag\revadd{, although no tracer emerges as an independent driver when the environmental parameters are analysed jointly}. These trends have the same sign as in previous work, but are weaker in our sample, likely because of differences in sample selection, redshift range, and the heterogeneous origin of the light curves.

(vii) \textit{Residual post-standardization \revdel{signals}\revadd{trends} remain weak.} Although most of the environmental dependence is removed by the stretch correction, weak residual trends persist in tracers linked to the recent star-formation history and the older stellar component. This is consistent with \revdel{the dominant environmental pathway being the age--stretch relation}\revadd{a dominant empirical association between stretch and a covariant young-stellar-population axis}, while any additional dependence of standardized luminosity on the local environment is comparatively subtle.

(viii) \textit{\revdel{Two environmental components.}\revadd{A possible second environmental component.}} Taken together, the results point most clearly to \revdel{a primary age--stretch component}\revadd{a dominant empirical association between stretch and a covariant young-stellar-population axis}, best traced by the joint UV+optical luminosity-weighted age and largely absorbed by the $x_1$ standardization\revadd{, rather than uniquely identifying stellar age as the underlying parameter}. The evidence for a second, residual luminosity dependence is weaker, but appears most plausibly in the mass-weighted age and local sSFR tracers, with amplitudes of order $\sim 0.05$ mag. \revadd{We emphasise that this second component remains below $3\sigma$ and that several correlated tracers were examined, so it should be regarded as a plausible scenario to be tested with larger samples rather than a firm detection.}

Beyond the individual measurements, this work delivers a homogeneous analysis framework in which UV photometry and optical spectroscopy are matched at a common 1~kpc physical scale. In practice, the results show that\revadd{, within our sample,} UV imaging at the SN site is \revdel{the most efficient single observable for predicting light-curve stretch and recovering accurate local stellar ages}\revadd{among the observables most strongly associated with the light-curve stretch, and it tightens the inferred local stellar ages within the adopted modelling framework}, while the combination of UV and optical diagnostics remains essential for characterizing the weaker residual trends that persist after standardization. Although HST-quality UV imaging is restricted to nearby hosts, several next-generation facilities will access the rest-frame F275W regime at intermediate and high redshift. From the ground, LSST $ugr$ imaging will sample rest-frame F275W at $z \gtrsim 0.3$, $0.75$, and $1.3$, respectively. From space, the bluest filters of the Nancy Grace Roman Space Telescope Wide Field Instrument (WFI; \citealt{2024SPIE13092E..0SS}) in the High-Latitude Time-Domain Survey (HLTDS; \citealt{2021arXiv211103081R}), F062 and F087, will provide matching coverage at $z \gtrsim 1.3$ and $\gtrsim 2.2$, while the Lazuli Space Observatory \citep{2026arXiv260102556R}, combining the Wide-field Context Camera (WCC) and the Integral Field Spectrograph, will access rest-frame F275W at $z \gtrsim 0.45$, providing a UV+IFS analogue of the present dataset directly from space. Together, these facilities will extend the framework established here from the local Universe to $z \sim 1$--$2$, enabling environmental-systematics studies for the next generation of SN~Ia cosmology. Finally, the age resolution in the tens-to-hundreds of Myr regime delivered by the joint UV+IFS spectrophotometric fits \revdel{enables a direct empirical determination of}\revadd{will help constrain} the SN~Ia delay-time distribution and its turn-on timescale\revadd{ (a full recovery additionally requires modelling the local star-formation histories, the survey selection function, and the SN detection efficiency)}, which will be presented in a forthcoming paper.

%%%%%%%%%%%%%%%%%%%%%%%%%%%%%%%%%%%%%%%%%%%%%%%%%%%%%%%
%%%%%%%%%%%%%%%%%%%%%%%%%%%%%%%%%%%%%%%%%%%%%%%%%%%%%%%
%%%%%%%%%%%%%%%%%%%%%%%%%%%%%%%%%%%%%%%%%%%%%%%%%%%%%%%

\begin{acknowledgements}
%L.G., A.A., E.D.G., A.G.S., C.P.G., and R.S 
We thank the anonymous referee for raising a few points that have significantly improved the quality of the paper. 
Th SNICE group acknowledges financial support from CSIC, MCIN and AEI 10.13039/501100011033 under projects PID2023-151307NB-I00, PIE 20215AT016, and CEX2020-001058-M. 
C.A. and C.B. acknowledge support from grants HST-GO-16287, HST-GO-16741, and HST-GO-17179 from Space Telescope Science Institute.
M.D. Stritzinger is funded by the Independent Research Fund Denmark (IRFD, grant number 10.46540/2032-00022B). 
Based on observations made with ESO Telescopes at the La Silla Paranal Observatory under programmes ID 60.A-9100, 094.B-0241, 094.B-0298, 094.B-0592, 094.B-0733, 095.B-0482, 095.B-0532, 095.B-0624, 095.D-0091, 095.D-0091, 096.B-0230, 096.D-0263, 096.D-0296, 097.A-0366, 097.A-0366, 097.B-0640, 097.D-0408, 098.D-0115, 099.A-0870, 099.B-0137, 099.B-0193, 099.B-0242, 099.B-0294, 099.D-0022, 0100.D-0341, 0101.B-0706, 0101.D-0748, 0102.D-0095, 0103.A-0637, 0103.B-0450, 196.B-0578, and 296.B-5054.
This research is based on observations made with the NASA/ESA Hubble Space Telescope obtained from the Space Telescope Science Institute, which is operated by the Association of Universities for Research in Astronomy, Inc., under NASA contract NAS 5–26555. These observations are associated with program(s) SNAP 16741 \& 17179.
\end{acknowledgements}

\bibliographystyle{bibtex/aa}
\bibliography{bibtex/references}

\appendix

\begin{onecolumn}
    
\section{Additional tables}

% Requires \usepackage{rotating} and \newcommand{\nodata}{\ldots}
\begin{table*}[h]
\centering
\caption{Summary of the local MUSE environment parameters. Columns are grouped into stellar-population quantities from \texttt{STARLIGHT} and gas-phase quantities from the emission-line analysis. The H$\alpha$ SFR surface density and sSFR use the Chabrier-scaled \citet{KennicuttEvans2012} calibration. The full table is available electronically; only the first 40 rows are shown here.}
\label{tab:muse_summary}
\scriptsize
\setlength{\tabcolsep}{4pt}
\resizebox{\textwidth}{!}{
\begin{tabular}{lc|cccccc|ccccc}
\hline\hline
SN & $z_{\rm obs}$ & \multicolumn{6}{c|}{Stellar population} & \multicolumn{5}{c}{Gas phase} \\
\hline
 &  & $\log M_\star$ & $\langle \log t_\star \rangle_L$ & $\langle \log t_\star \rangle_M$ & $Z_{\star,L}$ & $Z_{\star,M}$ & $A_V^\star$ & $E(B-V)_{\rm gas}$ & ${\rm EW}({\rm H}\alpha)$ & $12+\log({\rm O/H})_{\rm O3N2}$ & $\log \Sigma_{\rm SFR,H\alpha}$ & $\log {\rm sSFR}$ \\
 &  & [dex] & [yr] & [yr] &  &  & [mag] & [mag] & [\AA] & [dex] & [$M_\odot\,{\rm yr}^{-1}\,{\rm kpc}^{-2}$] & [yr$^{-1}$] \\
\hline
ASAS14lq & 0.02617 & $6.555\pm0.032$ & $8.353\pm0.065$ & $8.400\pm0.039$ & $0.016\pm0.002$ & $0.018\pm0.001$ & $0.000\pm0.000$ & $0.000\pm0.138$ & $5.682\pm1.664$ & $8.630\pm0.184$ & $-4.100\pm0.224$ & $-10.158\pm0.224$ \\
ASAS14lw & 0.02089 & $5.556\pm0.004$ & $6.156\pm0.061$ & $6.151\pm0.060$ & $0.011\pm0.003$ & $0.011\pm0.003$ & $0.000\pm0.000$ & $0.000\pm0.000$ & $6.428\pm1.347$ & $8.539\pm0.146$ & $-4.781\pm0.143$ & $-9.840\pm0.143$ \\
ASASSN-13ch & 0.01647 & $7.553\pm0.008$ & $8.847\pm0.025$ & $8.851\pm0.025$ & $0.015\pm0.003$ & $0.015\pm0.003$ & $0.000\pm0.000$ & $0.000\pm0.000$ & $17.496\pm0.092$ & $8.450\pm0.008$ & $-2.976\pm0.002$ & $-10.032\pm0.002$ \\
ASASSN-13cj & 0.01646 & $9.552\pm0.019$ & $9.538\pm0.046$ & $9.957\pm0.029$ & $0.031\pm0.003$ & $0.024\pm0.001$ & $0.093\pm0.024$ & $0.000\pm0.000$ & $1.523\pm0.009$ & $8.523\pm0.004$ & $-2.781\pm0.003$ & $-11.836\pm0.003$ \\
ASASSN-13cp & 0.02351 & $8.247\pm0.032$ & $9.174\pm0.027$ & $9.185\pm0.046$ & $0.050\pm0.000$ & $0.050\pm0.001$ & $0.019\pm0.015$ & $0.312\pm0.090$ & $4.490\pm0.046$ & $8.602\pm0.016$ & $-2.957\pm0.085$ & $-10.707\pm0.085$ \\
ASASSN-13cu & 0.02718 & $8.258\pm0.010$ & $8.752\pm0.035$ & $8.998\pm0.026$ & $0.020\pm0.001$ & $0.024\pm0.002$ & $0.150\pm0.052$ & $0.097\pm0.007$ & $45.320\pm0.062$ & $8.629\pm0.002$ & $-1.873\pm0.007$ & $-9.634\pm0.007$ \\
ASASSN-14ba & 0.03265 & $9.005\pm0.035$ & $8.658\pm0.033$ & $9.480\pm0.073$ & $0.023\pm0.001$ & $0.023\pm0.001$ & $0.710\pm0.037$ & $0.272\pm0.006$ & $48.680\pm0.050$ & $8.678\pm0.002$ & $-1.324\pm0.006$ & $-9.832\pm0.006$ \\
ASASSN-14cu & 0.02473 & $9.162\pm0.051$ & $9.232\pm0.022$ & $9.406\pm0.126$ & $0.049\pm0.001$ & $0.045\pm0.005$ & $0.000\pm0.000$ & $0.000\pm0.000$ & $0.616\pm0.025$ & $8.677\pm0.015$ & $-3.127\pm0.018$ & $-11.792\pm0.018$ \\
ASASSN-14db & 0.03738 & $9.199\pm0.015$ & $9.376\pm0.024$ & $9.886\pm0.020$ & $0.019\pm0.001$ & $0.018\pm0.002$ & $0.036\pm0.016$ & $0.206\pm0.009$ & $19.391\pm0.033$ & $8.733\pm0.003$ & $-1.694\pm0.009$ & $-10.396\pm0.009$ \\
ASASSN-14dd & 0.01784 & $7.399\pm0.006$ & $8.457\pm0.000$ & $8.457\pm0.000$ & $0.023\pm0.004$ & $0.023\pm0.004$ & $0.000\pm0.000$ & $0.000\pm0.000$ & $16.438\pm0.130$ & $8.732\pm0.016$ & $-2.955\pm0.004$ & $-9.856\pm0.004$ \\
ASASSN-14dz & 0.02232 & $9.240\pm0.018$ & $9.643\pm0.019$ & $9.999\pm0.009$ & $0.020\pm0.001$ & $0.020\pm0.000$ & $0.021\pm0.010$ & $0.000\pm0.000$ & $0.394\pm0.019$ & $8.607\pm0.027$ & $-3.726\pm0.023$ & $-12.469\pm0.023$ \\
ASASSN-14fa & 0.02232 & $7.945\pm0.029$ & $9.280\pm0.063$ & $9.967\pm0.028$ & $0.041\pm0.002$ & $0.035\pm0.003$ & $0.130\pm0.031$ & $0.000\pm0.000$ & $0.913\pm0.108$ & $8.612\pm0.094$ & $-4.539\pm0.049$ & $-11.987\pm0.049$ \\
ASASSN-14hr & 0.03355 & $8.689\pm0.017$ & $9.334\pm0.029$ & $9.763\pm0.030$ & $0.016\pm0.002$ & $0.017\pm0.001$ & $0.012\pm0.015$ & $0.102\pm0.031$ & $9.414\pm0.047$ & $8.769\pm0.011$ & $-2.548\pm0.029$ & $-10.740\pm0.029$ \\
ASASSN-14hu & 0.02163 & $6.994\pm0.012$ & $8.457\pm0.000$ & $8.457\pm0.000$ & $0.038\pm0.009$ & $0.039\pm0.008$ & $0.000\pm0.000$ & $0.000\pm0.000$ & $1.936\pm0.578$ & $8.691\pm0.105$ & $-3.942\pm0.068$ & $-10.438\pm0.068$ \\
ASASSN-14ig & 0.02911 & $9.613\pm0.035$ & $9.702\pm0.057$ & $9.959\pm0.034$ & $0.011\pm0.003$ & $0.017\pm0.002$ & $0.550\pm0.070$ & $0.002\pm0.016$ & $1.486\pm0.009$ & $8.716\pm0.006$ & $-2.832\pm0.016$ & $-11.948\pm0.016$ \\
ASASSN-14jc & 0.01112 & $8.851\pm0.048$ & $9.082\pm0.028$ & $9.492\pm0.093$ & $0.020\pm0.002$ & $0.020\pm0.002$ & $0.190\pm0.030$ & $0.004\pm0.007$ & $15.264\pm0.023$ & $8.761\pm0.004$ & $-2.135\pm0.006$ & $-10.489\pm0.006$ \\
ASASSN-14jg & 0.01481 & $7.435\pm0.052$ & $7.913\pm0.036$ & $8.337\pm0.081$ & $0.016\pm0.001$ & $0.019\pm0.003$ & $0.690\pm0.040$ & $0.139\pm0.004$ & $71.796\pm0.073$ & $8.530\pm0.001$ & $-2.222\pm0.004$ & $-9.160\pm0.004$ \\
ASASSN-14kq & 0.03375 & $7.349\pm0.013$ & $8.739\pm0.026$ & $8.778\pm0.032$ & $0.016\pm0.003$ & $0.016\pm0.003$ & $0.000\pm0.001$ & $0.000\pm0.000$ & $18.030\pm0.324$ & $8.495\pm0.021$ & $-3.129\pm0.010$ & $-9.981\pm0.010$ \\
ASASSN-14lt & 0.03189 & $9.061\pm0.034$ & $9.720\pm0.059$ & $9.945\pm0.032$ & $0.022\pm0.006$ & $0.018\pm0.001$ & $0.000\pm0.000$ & $0.000\pm0.149$ & $0.209\pm0.020$ & $8.599\pm0.058$ & $-4.077\pm0.163$ & $-12.641\pm0.163$ \\
ASASSN-14lv & 0.04928 & $8.416\pm0.036$ & $9.119\pm0.033$ & $9.436\pm0.093$ & $0.021\pm0.002$ & $0.020\pm0.001$ & $0.080\pm0.019$ & $0.268\pm0.029$ & $11.187\pm0.082$ & $8.683\pm0.011$ & $-2.410\pm0.028$ & $-10.328\pm0.028$ \\
\hline
\end{tabular}}
\end{table*}

% Requires \usepackage{rotating} and \newcommand{\nodata}{\ldots}
\begin{table*}[h]
\centering
\caption{Summary of the local UV and optical photometric parameters. The photometric quantities are corrected for Milky Way foreground extinction only, while the host-galaxy attenuation terms derived from the local joint UV+optical 	exttt{STARLIGHT} fit are reported separately when available. The F275W$-r$ color and derived UV quantities are corrected for host attenuation when available, and otherwise use the Milky Way-corrected values. The full table is available electronically; only the first 40 rows are shown here.}
\label{tab:photometry_summary}
\scriptsize
\setlength{\tabcolsep}{4pt}
\resizebox{\textwidth}{!}{
\begin{tabular}{l|cc|ccc|c|cc|l}
\hline\hline
SN & \multicolumn{2}{c|}{MW-corrected photometry} & \multicolumn{3}{c|}{Host attenuation} & \multicolumn{1}{c|}{Host-corrected color} & \multicolumn{2}{c|}{Derived UV quantities} & $r$ survey \\
\hline
 & $M_{\rm F275W}$ & $M_r$ & $E(B-V)_\star$ & $A_{\rm F275W}$ & $A_r$ & ${\rm F275W}-r$ & $\log \Sigma_{\rm UV}$ & $\log \Sigma_{\rm SFR,UV}$ &  \\
 & [mag] & [mag] & [mag] & [mag] & [mag] & [mag] & [erg s$^{-1}$ Hz$^{-1}$ kpc$^{-2}$] & [$M_\odot\,{\rm yr}^{-1}\,{\rm kpc}^{-2}$] &  \\
\hline
ASAS14lq & $-11.992\pm0.267$ & $>-12.060$ & $0.000\pm0.000$ & $0.000\pm0.000$ & $0.000\pm0.000$ & $<0.067$ & $24.938\pm0.108$ & $-3.029\pm0.108$ & PanSTARRS \\
ASAS14lw & $-13.780\pm0.038$ & $-11.861\pm0.679$ & $0.000\pm0.000$ & $0.000\pm0.000$ & $0.000\pm0.000$ & $-1.918\pm0.680$ & $25.653\pm0.015$ & $-2.314\pm0.015$ & SkyMapper \\
ASASSN-13ch & $-12.584\pm0.119$ & $-15.355\pm0.157$ & $0.000\pm0.000$ & $0.000\pm0.000$ & $0.000\pm0.000$ & $2.771\pm0.195$ & $25.174\pm0.046$ & $-2.792\pm0.046$ & LegacySurvey \\
ASASSN-13cj & $-14.135\pm0.023$ & $-18.386\pm0.033$ & $0.035\pm0.008$ & $0.221\pm0.053$ & $0.090\pm0.022$ & $4.140\pm0.068$ & $25.870\pm0.022$ & $-2.097\pm0.022$ & LegacySurvey \\
ASASSN-13cp & $>-11.445$ & $-16.292\pm0.308$ & $0.004\pm0.004$ & $0.026\pm0.027$ & $0.011\pm0.011$ & $>4.807$ & $<24.735$ & $<-3.232$ & LegacySurvey \\
ASASSN-13cu & $-14.689\pm0.021$ & $-17.008\pm0.116$ & $0.053\pm0.017$ & $0.334\pm0.106$ & $0.136\pm0.043$ & $2.140\pm0.162$ & $26.137\pm0.042$ & $-1.829\pm0.042$ & LegacySurvey \\
ASASSN-14ba & $-15.246\pm0.015$ & $-17.735\pm0.014$ & $0.227\pm0.012$ & $1.421\pm0.074$ & $0.580\pm0.030$ & $1.642\pm0.081$ & $26.812\pm0.030$ & $-1.155\pm0.030$ & LegacySurvey \\
ASASSN-14cu & $-13.665\pm0.060$ & $-18.417\pm0.075$ & $0.000\pm0.000$ & $0.000\pm0.000$ & $0.000\pm0.000$ & $4.752\pm0.095$ & $25.607\pm0.024$ & $-2.360\pm0.024$ & LegacySurvey \\
ASASSN-14db & $-14.934\pm0.025$ & $-18.006\pm0.248$ & $0.017\pm0.006$ & $0.105\pm0.036$ & $0.043\pm0.015$ & $3.030\pm0.253$ & $26.143\pm0.017$ & $-1.824\pm0.017$ & LegacySurvey \\
ASASSN-14dd & $-13.579\pm0.063$ & $-15.279\pm0.062$ & $0.000\pm0.000$ & $0.000\pm0.000$ & $0.000\pm0.000$ & $1.700\pm0.087$ & $25.572\pm0.024$ & $-2.394\pm0.024$ & SkyMapper \\
ASASSN-14dz & $-13.458\pm0.049$ & $-17.726\pm0.062$ & $0.007\pm0.003$ & $0.044\pm0.020$ & $0.018\pm0.008$ & $4.243\pm0.084$ & $25.541\pm0.022$ & $-2.426\pm0.022$ & LegacySurvey \\
ASASSN-14fa & $-11.966\pm0.176$ & $-14.682\pm0.019$ & $0.038\pm0.010$ & $0.241\pm0.061$ & $0.098\pm0.025$ & $2.560\pm0.191$ & $25.032\pm0.075$ & $-2.934\pm0.075$ & DES \\
ASASSN-14hr & $-14.138\pm0.037$ & $-16.938\pm0.007$ & $0.005\pm0.005$ & $0.031\pm0.029$ & $0.013\pm0.012$ & $2.786\pm0.049$ & $25.805\pm0.019$ & $-2.161\pm0.019$ & DES \\
ASASSN-14hu & $-12.253\pm0.166$ & $-14.017\pm0.118$ & $0.000\pm0.000$ & $0.000\pm0.000$ & $0.000\pm0.000$ & $1.765\pm0.208$ & $25.042\pm0.069$ & $-2.925\pm0.069$ & LegacySurvey \\
ASASSN-14ig & $-13.551\pm0.055$ & $-18.264\pm0.007$ & $0.156\pm0.025$ & $0.974\pm0.155$ & $0.398\pm0.063$ & $4.056\pm0.175$ & $26.005\pm0.065$ & $-1.962\pm0.065$ & DES \\
ASASSN-14jc & $-15.117\pm0.010$ & $-17.570\pm0.010$ & $0.057\pm0.010$ & $0.358\pm0.061$ & $0.146\pm0.025$ & $2.247\pm0.068$ & $26.341\pm0.025$ & $-1.626\pm0.025$ & SkyMapper \\
ASASSN-14jg & $-14.103\pm0.017$ & $-16.986\pm0.006$ & $0.222\pm0.013$ & $1.390\pm0.079$ & $0.568\pm0.032$ & $2.060\pm0.086$ & $26.339\pm0.032$ & $-1.628\pm0.032$ & DES \\
ASASSN-14kq & $-12.527\pm0.167$ & $-14.818\pm0.060$ & $0.000\pm0.000$ & $0.000\pm0.002$ & $0.000\pm0.001$ & $2.291\pm0.184$ & $25.152\pm0.069$ & $-2.815\pm0.069$ & PanSTARRS \\
ASASSN-14lt & $-12.799\pm0.144$ & $-17.970\pm0.226$ & $0.000\pm0.000$ & $0.000\pm0.000$ & $0.000\pm0.000$ & $5.172\pm0.266$ & $25.260\pm0.055$ & $-2.706\pm0.055$ & LegacySurvey \\
ASASSN-14lv & $-13.601\pm0.097$ & $-16.756\pm0.008$ & $0.028\pm0.006$ & $0.178\pm0.039$ & $0.073\pm0.016$ & $3.060\pm0.103$ & $25.645\pm0.040$ & $-2.321\pm0.040$ & DES \\
\hline
\end{tabular}}
\end{table*}

% Requires \newcommand{\nodata}{\ldots}
\begin{table*}[h]
\centering
\caption{Summary of the \texttt{sncosmo} SN light-curve parameters and Hubble residuals. The full table is available electronically; only the first 20 rows are shown here.}
\label{tab:lc_hr_summary_sncosmo_only}
\scriptsize
\setlength{\tabcolsep}{4pt}
\resizebox{0.575\textwidth}{!}{
\begin{tabular}{lcc|cccc}
\hline\hline
SN & $z_{\rm obs}$ & $z_{\rm CMB}$ & \multicolumn{4}{c}{sncosmo} \\
\hline
 &  &  & $m_{B,\max}$ & $x_1$ & $c$ & HR \\
\hline
ASAS14lq & 0.02617 & 0.02480 & \nodata & \nodata & \nodata & \nodata \\
ASAS14lw & 0.02089 & 0.02014 & $15.702\pm0.001$ & $2.335\pm0.013$ & $-0.003\pm0.001$ & $0.355\pm0.003$ \\
ASASSN-13ch & 0.01647 & 0.01684 & \nodata & \nodata & \nodata & \nodata \\
ASASSN-13cj & 0.01646 & 0.01682 & \nodata & \nodata & \nodata & \nodata \\
ASASSN-13cp & 0.02351 & 0.02398 & \nodata & \nodata & \nodata & \nodata \\
ASASSN-13cu & 0.02718 & 0.02597 & \nodata & \nodata & \nodata & \nodata \\
ASASSN-14ba & 0.03265 & 0.03393 & \nodata & \nodata & \nodata & \nodata \\
ASASSN-14cu & 0.02473 & 0.02594 & \nodata & \nodata & \nodata & \nodata \\
ASASSN-14db & 0.03738 & 0.03717 & \nodata & \nodata & \nodata & \nodata \\
ASASSN-14dd & 0.01784 & 0.01825 & \nodata & \nodata & \nodata & \nodata \\
ASASSN-14dz & 0.02232 & 0.02297 & \nodata & \nodata & \nodata & \nodata \\
ASASSN-14fa & 0.02232 & 0.02172 & \nodata & \nodata & \nodata & \nodata \\
ASASSN-14hr & 0.03360 & 0.03255 & $17.211\pm0.003$ & $-1.692\pm0.021$ & $0.169\pm0.002$ & $-0.036\pm0.005$ \\
ASASSN-14hu & 0.02159 & 0.02195 & $15.516\pm0.002$ & $1.471\pm0.017$ & $-0.037\pm0.001$ & $-0.003\pm0.003$ \\
ASASSN-14ig & 0.02911 & 0.02869 & \nodata & \nodata & \nodata & \nodata \\
ASASSN-14jc & 0.01132 & 0.01161 & $15.680\pm0.003$ & $-0.664\pm0.016$ & $0.452\pm0.001$ & $0.011\pm0.005$ \\
ASASSN-14jg & 0.01482 & 0.01427 & $14.821\pm0.002$ & $1.835\pm0.018$ & $0.040\pm0.001$ & $0.069\pm0.004$ \\
ASASSN-14kq & 0.03358 & 0.03266 & $16.656\pm0.002$ & $0.789\pm0.027$ & $0.020\pm0.001$ & $0.038\pm0.005$ \\
ASASSN-14lt & 0.03202 & 0.03122 & $16.408\pm0.003$ & $-0.323\pm0.045$ & $-0.006\pm0.001$ & $-0.135\pm0.006$ \\
ASASSN-14lv & 0.04928 & 0.04836 & \nodata & \nodata & \nodata & \nodata \\
\hline
\end{tabular}}
\end{table*}

% Requires \usepackage{rotating} and \usepackage{graphicx}
\begin{table*}
\centering
\caption{Correlation and median-split summary for \texttt{sncosmo} parameters ($x_1$, $c$, and HR$_{SALT}$) versus local environment quantities, restricted to the 169-object Tripp-calibration sample. The two bins are split at the median of each x-variable; low/high bins report medians with robust standard errors (nMAD/$\sqrt{N}$). Difference (h-l) is the median offset with propagated uncertainty, and the last column reports the significance of this difference. For UV-derived quantities with limits, limit objects are included using their limit values both for the median x-split and for low/high-bin summaries.}
\label{tab:env_sncosmo_median_split_summary_calibration}
\scriptsize
\setlength{\tabcolsep}{4pt}
\resizebox{\textwidth}{!}{
\begin{tabular}{llrrrrlllr}
\hline\hline
y variable & x variable & N & Pearson r & p-value & x cut & low bin & high bin & Difference (h-l) & $\sigma$ \\
\hline
x1 & 12+$\log$ (O/H) & 163 & -0.239 & 0.002078 & 8.684 & 0.232 +- 0.135 & -0.437 +- 0.159 & -0.669 +- 0.209 & 3.21 \\
  & F275W-$r$ & 166 & -0.418 & 2.781e-06 & 2.151 & 0.601 +- 0.103 & -0.507 +- 0.138 & -1.108 +- 0.172 & 6.44 \\
  & $\log \Sigma$SFR$_{H\alpha}$ & 163 & 0.107 & 0.1727 & -3.145 & -0.382 +- 0.191 & 0.126 +- 0.112 & 0.509 +- 0.222 & 2.29 \\
  & $\log \Sigma$SFR$_{UV}$ & 169 & 0.251 & 0.005808 & -2.49 & -0.306 +- 0.161 & 0.309 +- 0.127 & 0.615 +- 0.205 & 3.00 \\
  & $\log$ EW$_{H_\alpha}$ & 162 & 0.397 & 1.692e-07 & 0.9167 & -0.539 +- 0.190 & 0.380 +- 0.102 & 0.920 +- 0.215 & 4.27 \\
  & $\log M$ & 169 & -0.361 & 1.425e-06 & 7.907 & 0.551 +- 0.130 & -0.395 +- 0.146 & -0.946 +- 0.196 & 4.84 \\
  & $\log Z_L$ & 169 & -0.112 & 0.1461 & 0.02498 & 0.325 +- 0.137 & -0.262 +- 0.144 & -0.587 +- 0.199 & 2.95 \\
  & $\log Z_M$ & 169 & -0.027 & 0.7272 & 0.02318 & 0.079 +- 0.150 & -0.025 +- 0.129 & -0.104 +- 0.198 & 0.52 \\
  & $\log$ sSFR$_{H\alpha}$ & 163 & 0.452 & 1.37e-09 & -10.38 & -0.817 +- 0.162 & 0.546 +- 0.106 & 1.363 +- 0.194 & 7.04 \\
  & $\log$ sSFR$_{UV}$ & 169 & 0.506 & 4.52e-09 & -9.713 & -0.638 +- 0.155 & 0.612 +- 0.097 & 1.250 +- 0.183 & 6.84 \\
  & $\log t_L$ & 169 & -0.475 & 6.805e-11 & 8.807 & 0.606 +- 0.098 & -0.762 +- 0.149 & -1.368 +- 0.179 & 7.66 \\
  & $\log t_M$ & 169 & -0.380 & 3.576e-07 & 9.227 & 0.401 +- 0.118 & -0.507 +- 0.164 & -0.909 +- 0.202 & 4.50 \\
\hline
c & 12+$\log$ (O/H) & 163 & 0.202 & 0.00954 & 8.684 & 0.003 +- 0.008 & 0.021 +- 0.010 & 0.018 +- 0.012 & 1.45 \\
  & F275W-$r$ & 166 & 0.093 & 0.3189 & 2.151 & -0.003 +- 0.009 & 0.025 +- 0.008 & 0.028 +- 0.012 & 2.35 \\
  & $\log \Sigma$SFR$_{H\alpha}$ & 163 & 0.178 & 0.02317 & -3.145 & 0.007 +- 0.007 & 0.013 +- 0.010 & 0.007 +- 0.012 & 0.53 \\
  & $\log \Sigma$SFR$_{UV}$ & 169 & 0.062 & 0.5018 & -2.49 & 0.019 +- 0.007 & -0.003 +- 0.009 & -0.022 +- 0.011 & 1.95 \\
  & $\log$ EW$_{H_\alpha}$ & 162 & -4.453e-04 & 0.9955 & 0.9167 & 0.017 +- 0.008 & 0.003 +- 0.010 & -0.014 +- 0.013 & 1.10 \\
  & $\log M$ & 169 & 0.273 & 0.0003278 & 7.907 & -0.009 +- 0.009 & 0.031 +- 0.009 & 0.040 +- 0.012 & 3.29 \\
  & $\log Z_L$ & 169 & -0.008 & 0.9213 & 0.02498 & 0.011 +- 0.009 & 0.003 +- 0.008 & -0.008 +- 0.012 & 0.66 \\
  & $\log Z_M$ & 169 & -0.035 & 0.6517 & 0.02318 & 0.008 +- 0.010 & 0.008 +- 0.007 & -0.001 +- 0.012 & 0.06 \\
  & $\log$ sSFR$_{H\alpha}$ & 163 & -0.056 & 0.4814 & -10.38 & 0.030 +- 0.009 & -0.006 +- 0.009 & -0.036 +- 0.012 & 2.91 \\
  & $\log$ sSFR$_{UV}$ & 169 & -0.123 & 0.1813 & -9.713 & 0.030 +- 0.009 & -0.004 +- 0.009 & -0.035 +- 0.012 & 2.83 \\
  & $\log t_L$ & 169 & 0.251 & 0.0009854 & 8.807 & -0.005 +- 0.008 & 0.030 +- 0.008 & 0.035 +- 0.012 & 2.99 \\
  & $\log t_M$ & 169 & 0.255 & 0.0008223 & 9.227 & -0.008 +- 0.008 & 0.031 +- 0.008 & 0.038 +- 0.012 & 3.23 \\
\hline
HR$_{SALT}$ & 12+$\log$ (O/H) & 163 & -0.089 & 0.2599 & 8.684 & 0.008 +- 0.016 & -0.015 +- 0.018 & -0.023 +- 0.024 & 0.96 \\
  & F275W-$r$ & 166 & -0.185 & 0.04538 & 2.151 & 0.008 +- 0.014 & -0.010 +- 0.018 & -0.018 +- 0.023 & 0.82 \\
  & $\log \Sigma$SFR$_{H\alpha}$ & 163 & -0.014 & 0.8601 & -3.145 & -0.003 +- 0.018 & 0.004 +- 0.017 & 0.007 +- 0.025 & 0.29 \\
  & $\log \Sigma$SFR$_{UV}$ & 169 & 0.046 & 0.6173 & -2.49 & -0.011 +- 0.014 & 0.010 +- 0.018 & 0.021 +- 0.023 & 0.90 \\
  & $\log$ EW$_{H_\alpha}$ & 162 & 0.039 & 0.626 & 0.9167 & -0.025 +- 0.020 & 0.017 +- 0.014 & 0.042 +- 0.024 & 1.73 \\
  & $\log M$ & 169 & -0.085 & 0.2744 & 7.907 & 0.009 +- 0.014 & -0.035 +- 0.018 & -0.044 +- 0.023 & 1.91 \\
  & $\log Z_L$ & 169 & -0.049 & 0.5293 & 0.02498 & 0.017 +- 0.015 & -0.019 +- 0.018 & -0.036 +- 0.023 & 1.55 \\
  & $\log Z_M$ & 169 & -0.072 & 0.3523 & 0.02318 & 0.029 +- 0.017 & -0.021 +- 0.015 & -0.050 +- 0.022 & 2.23 \\
  & $\log$ sSFR$_{H\alpha}$ & 163 & 0.060 & 0.4448 & -10.38 & -0.025 +- 0.020 & 0.025 +- 0.014 & 0.050 +- 0.024 & 2.06 \\
  & $\log$ sSFR$_{UV}$ & 169 & 0.174 & 0.05819 & -9.713 & -0.035 +- 0.018 & 0.010 +- 0.012 & 0.045 +- 0.022 & 2.09 \\
  & $\log t_L$ & 169 & -0.112 & 0.1461 & 8.807 & 0.009 +- 0.013 & -0.026 +- 0.018 & -0.035 +- 0.022 & 1.59 \\
  & $\log t_M$ & 169 & -0.160 & 0.03715 & 9.227 & 0.019 +- 0.014 & -0.036 +- 0.017 & -0.055 +- 0.022 & 2.49 \\
\hline
\end{tabular}%
}
\end{table*}

\clearpage

\section{Additional figures} \label{app:figs}

\begin{figure*}[h]
\centering
\includegraphics[width=0.95\textwidth]{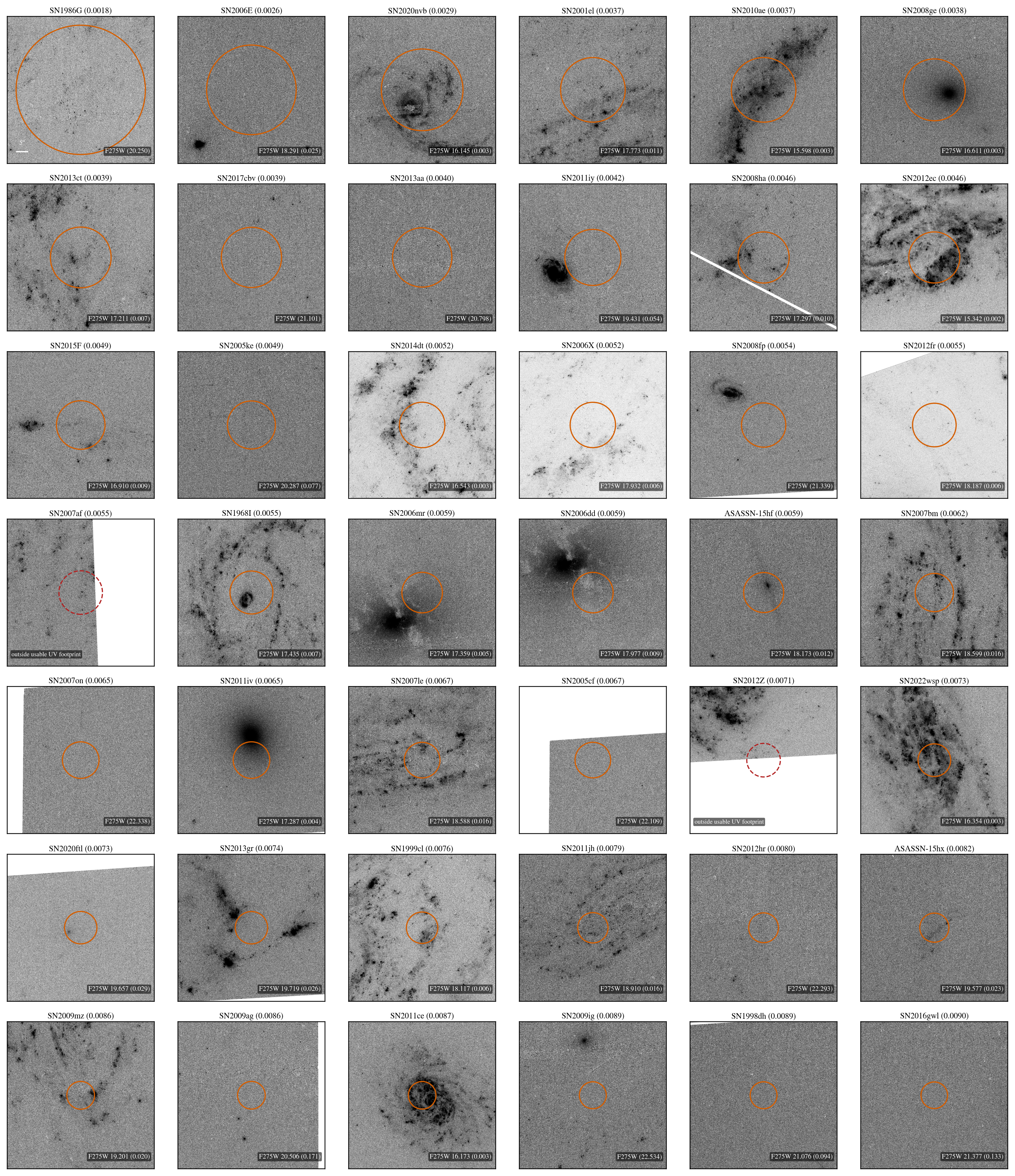}
\caption{HST/WFC3 F275W cutouts for the first 42 SN environments in order of increasing redshift. Each panel shows a common $60\arcsec \times 60\arcsec$ field centered on the SN position, with the orange circle marking the projected 1~kpc-radius aperture used throughout the analysis. Panel titles list the SN name and redshift. The lower-right annotation reports the measured F275W magnitude or upper limit where available. For SN~2007af and SN~2012Z the dashed red aperture indicates that the nominal 1~kpc aperture extends outside the usable HST footprint, so those locations are excluded from the UV photometric analysis.}
\label{fig:hst_uv_appendix_grid_first42}
\end{figure*}

\begin{figure*}[h]
\centering
\includegraphics[width=0.86\textwidth]{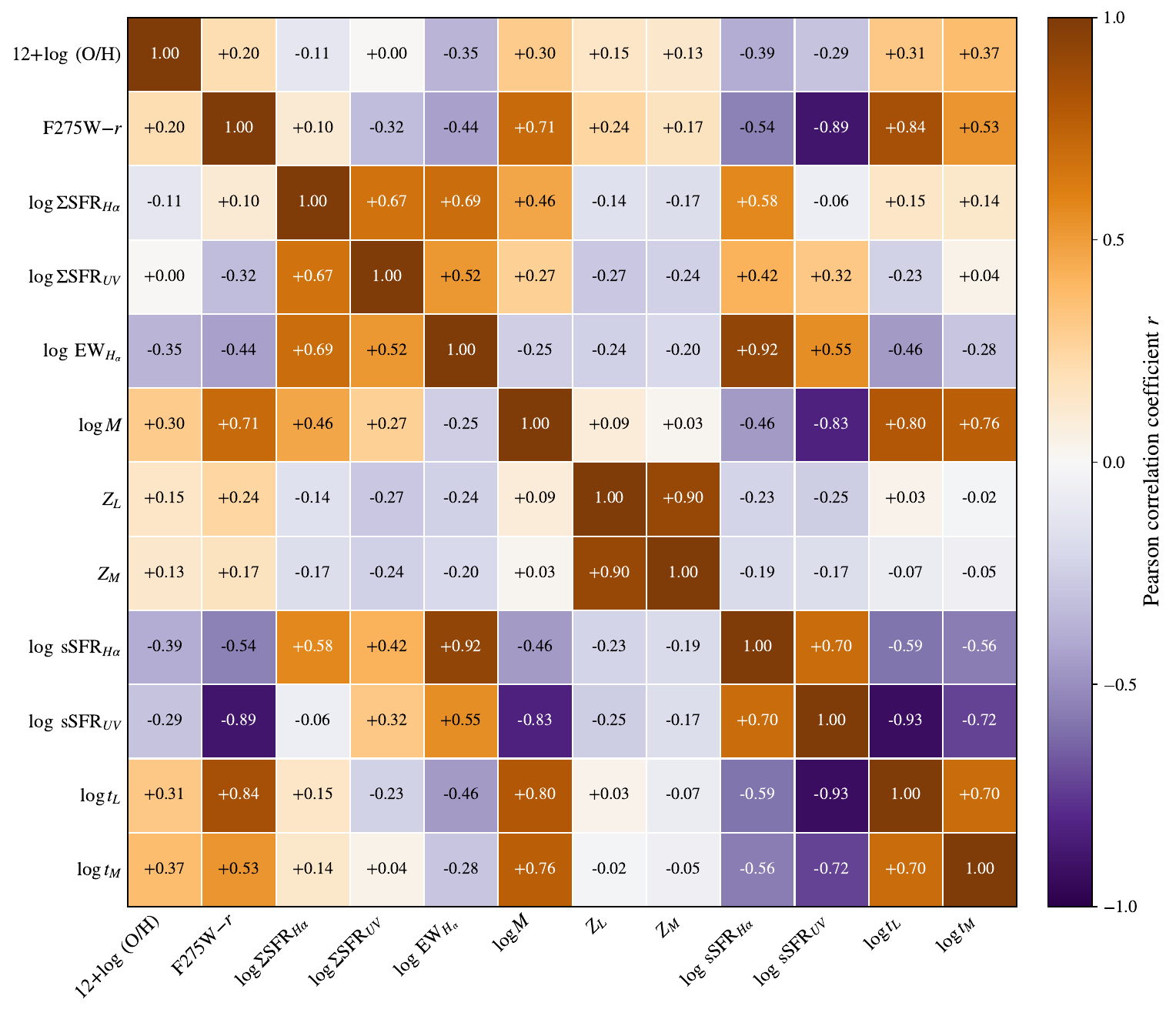}
\caption{Correlation matrix of the twelve local environmental tracers on the 169-object calibration sample (complete-case subset with all tracers detected, $N=117$). Each cell gives the Pearson correlation coefficient $r$ between the pair of tracers labelled on the two axes, and is color-coded on the diverging scale at right, from $r=-1$ (purple, perfect anticorrelation) through $r=0$ (white, no correlation) to $r=+1$ (orange, perfect correlation). }
\label{fig:tracer_corr}
\end{figure*}

\clearpage

\section{Sample representativeness}\label{app:repr}

\revadd{The target selection is driven by the joint availability of HST/F275W imaging, MUSE IFS, and well-sampled SN~Ia light curves, and is therefore not volume-limited. To assess how this selection propagates through the analysis, we compare the parent (340), fitted (211), and Tripp-calibration (169) samples. Figure~\ref{fig:point5_cdf} shows the cumulative distributions of the local stellar mass, the luminosity- and mass-weighted ages, and the local H$\alpha$- and UV-based specific SFRs. Table~\ref{tab:point5_dist} lists the sample medians, the effective number of measurements, and the nested two-sample Kolmogorov--Smirnov (KS) and Anderson--Darling (AD) $p$-values for each pair. Because the three samples are nested, these pairwise KS and AD tests are not independent and are used only descriptively. This motivates the \emph{independent} comparison in Table~\ref{tab:point5_excl} between the retained calibration sample and the excluded remainder. Table~\ref{tab:point5_frac} gives the fractions of UV-detected, targeted, and star-forming environments with 68\% binomial confidence intervals. Discovery strategy is assigned by the construction of the sample: SNe from wide-field untargeted surveys (La\,Silla-QUEST, ASAS-SN, iPTF, Pan-STARRS, Catalina, KISS and OGLE) together with the CSP-II follow-up sample, which by design targeted transients found by untargeted surveys, form the untargeted class (210/340), while the CSP-I follow-up of galaxy-targeted and amateur discoveries forms the targeted class (130/340). The full composition is given in Table~\ref{tab:point5_disc}.}

\revadd{The nested distributions are broadly similar, although the local stellar mass differs formally between the parent and calibration samples (KS $p=0.05$, AD $p=0.018$), so the two are not fully consistent for that quantity. The light-curve and cosmology cuts preferentially retain lower-mass local environments ($\log M_\star$ median $7.91$ versus $8.41$; KS $p<10^{-3}$), together with younger luminosity-weighted ages ($8.81$ versus $8.99$; KS $p=0.008$) and higher UV sSFR ($-9.69$ versus $-10.13$; KS $p=0.009$), while the H$\alpha$ sSFR shows no significant shift. The evidence for a mass-weighted-age shift is test-dependent, being not significant under KS ($p=0.097$) but marginally significant under AD ($p=0.019$). The UV-detected, targeted, and star-forming fractions are stable across the three samples, with all pairwise differences below $1\sigma$. This mild bias toward lower-mass, younger, more star-forming sites in the calibration sample is consistent with the diluted Hubble-residual steps discussed in Sect.~\ref{sec:hr}.}

\begin{figure*}[h]
\centering
\includegraphics[width=\textwidth]{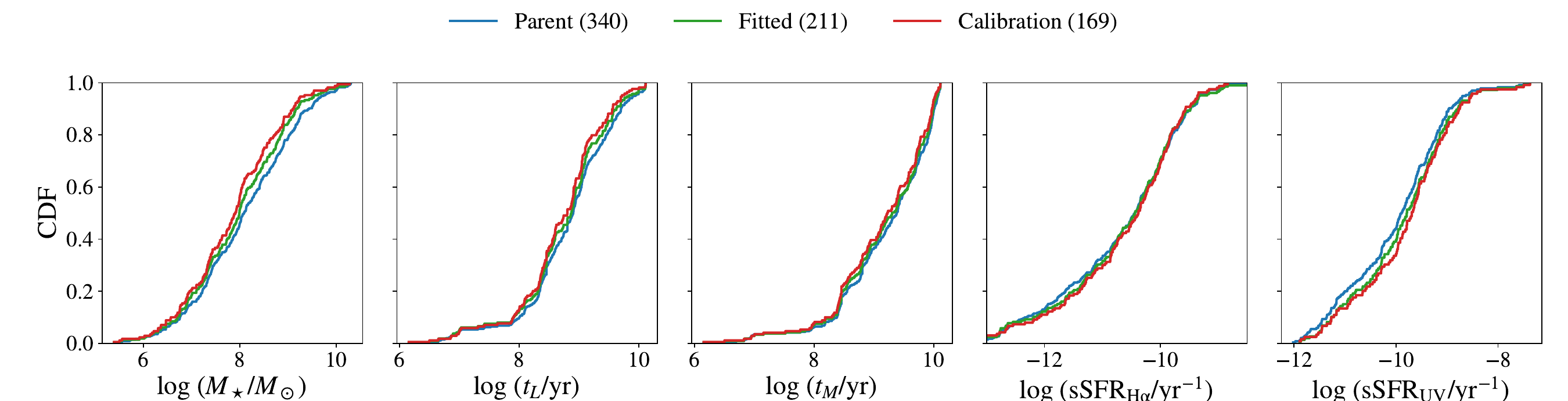}
\caption{Cumulative distributions of local environmental properties for the parent (340; blue), fitted (211; green) and Tripp-calibration (169; red) samples: local stellar mass, luminosity- and mass-weighted stellar age, and local H$\alpha$- and UV-based specific SFR.}
\label{fig:point5_cdf}
\end{figure*}

\begin{table*}[h]
\centering
\caption{Local environmental distributions across the parent (340), fitted (211) and calibration (169) samples: medians, with the number of detected measurements $N$ in parentheses (UV upper limits excluded), and two-sample KS\,/\,AD $p$-values for each nested pair (P, F, C). AD $p$-values are obtained by permutation ($5000$ resamples).}
\label{tab:point5_dist}
\begin{tabular}{lcccccc}
\hline\hline
Quantity & med$_{340}$\,($N$) & med$_{211}$\,($N$) & med$_{169}$\,($N$) & P--F & P--C & F--C \\
 & & & & \multicolumn{3}{c}{KS\,/\,AD $p$} \\
\hline
$\log M_\star$ & 8.09\,(337) & 7.99\,(211) & 7.91\,(169) & $0.546/0.229$ & $0.052/0.018$ & $0.799/0.642$ \\
$\log t_L$ & 8.90\,(337) & 8.87\,(211) & 8.81\,(169) & $0.789/0.579$ & $0.305/0.099$ & $0.967/0.853$ \\
$\log t_M$ & 9.38\,(337) & 9.30\,(211) & 9.23\,(169) & $0.993/0.918$ & $0.683/0.292$ & $0.998/0.960$ \\
$\log\,\mathrm{sSFR}_{\rm H\alpha}$ & -10.44\,(330) & -10.44\,(205) & -10.38\,(163) & $0.998/0.997$ & $0.958/0.891$ & $1.000/0.999$ \\
$\log\,\mathrm{sSFR}_{\rm UV}$ & -9.89\,(235) & -9.79\,(146) & -9.69\,(119) & $0.792/0.433$ & $0.336/0.111$ & $0.967/0.976$ \\
\hline
\end{tabular}
\end{table*}

\begin{table}[h]
\centering
\caption{Independent retained-versus-excluded comparison: the Tripp-calibration sample (169) against the remaining 171 objects of the parent sample. Medians, number of measurements in each subset, and two-sample KS\,/\,AD $p$-values (AD by permutation).}
\label{tab:point5_excl}
\begin{tabular}{lcccc}
\hline\hline
Quantity & med$_{\rm ret}$ & med$_{\rm exc}$ & $N_{\rm ret}/N_{\rm exc}$ & KS\,/\,AD $p$ \\
\hline
local $\log M_\star$ & 7.91 & 8.41 & 169/168 & ${<}0.001/0.0002$ \\
$\log t_L$ & 8.81 & 8.99 & 169/168 & $0.008/0.0018$ \\
$\log t_M$ & 9.23 & 9.44 & 169/168 & $0.097/0.0198$ \\
$\log\,\mathrm{sSFR}_{\rm H\alpha}$ & -10.38 & -10.50 & 163/167 & $0.442/0.342$ \\
$\log\,\mathrm{sSFR}_{\rm UV}$ & -9.69 & -10.13 & 119/116 & $0.009/0.0012$ \\
\hline
\end{tabular}
\end{table}

\begin{table}[h]
\centering
\caption{Fractions of UV-detected, targeted, and star-forming environments in the three samples, with 68\% Wilson binomial intervals and $k/n$. The UV-detected fraction is evaluated over the environments with valid HST imaging (337, 209 and 169). ``Targeted'' denotes galaxy-targeted and amateur SN discoveries (Table~\ref{tab:point5_disc}).}
\label{tab:point5_frac}
\begin{tabular}{lccc}
\hline\hline
Fraction & Parent & Fitted & Calibration \\
\hline
UV-detected & $69.7^{+2.4}_{-2.6}$ (235/337) & $69.9^{+3.1}_{-3.3}$ (146/209) & $70.4^{+3.4}_{-3.6}$ (119/169) \\
Targeted vs untargeted & $38.2^{+2.7}_{-2.6}$ (130/340) & $41.7^{+3.4}_{-3.3}$ (88/211) & $36.7^{+3.8}_{-3.6}$ (62/169) \\
SF: $\log\,{\rm sSFR}_{\rm H\alpha}>-10.82$ & $62.1^{+2.6}_{-2.7}$ (205/330) & $63.4^{+3.3}_{-3.4}$ (130/205) & $64.4^{+3.7}_{-3.8}$ (105/163) \\
SF: EW(H$\alpha$)$>$3\,\AA & $65.9^{+2.6}_{-2.7}$ (216/328) & $68.1^{+3.2}_{-3.3}$ (139/204) & $69.1^{+3.5}_{-3.7}$ (112/162) \\
\hline
\end{tabular}
\end{table}

\revadd{\begin{table}[h]
\centering
\caption{Discovery-survey composition of the parent sample (340 SNe).}
\label{tab:point5_disc}
\begin{tabular}{lc}
\hline\hline
Discovery survey / programme & $N$ \\
\hline
La\,Silla-QUEST & 57 \\
ASAS-SN & 37 \\
(i)PTF & 15 \\
Pan-STARRS & 5 \\
Catalina (CRTS) & 3 \\
KISS & 3 \\
OGLE & 1 \\
CSP-II follow-up (untargeted, IAU-named) & 89 \\
\hline
Untargeted total & 210 \\
CSP-I / targeted-era (IAU/amateur) & 130 \\
\hline
Total & 340 \\
\hline
\end{tabular}
\end{table}}

\end{onecolumn}

\end{document}